\documentclass[prb,amsmath,amssymb,twocolumn]{revtex4-2}
\usepackage{graphicx}
\usepackage{subfigure}
\usepackage{adjustbox}
\usepackage{bm}
\usepackage{color}
\usepackage{braket}
\usepackage{standalone}
\usepackage{multirow}
\usepackage{tikz}
\usepackage{mathrsfs}
\usepackage{dsfont}
\usepackage{comment}
\usepackage{amsmath}
\usepackage[colorlinks,bookmarks=true,citecolor=blue,linkcolor=red,urlcolor=blue]{hyperref}
\usepackage{cleveref}

\newcommand{\be}{\begin{equation}}
\newcommand{\ee}{\end{equation}}

\newcommand{\ba}{\begin{eqnarray}}
\newcommand{\ea}{\end{eqnarray}}

\newcommand*{\id}{{\normalfont\hbox{1\kern-0.15em \vrule width .8pt depth-.5pt}}}

\graphicspath{{./figures/}}

\begin{document}

\title{Local density of states and particle entanglement in topological quantum fluids}

\author{Songyang Pu$^{1}$, Ajit C. Balram$^{2,3}$, and Zlatko Papi\'c$^{1}$}
\affiliation{$^{1}$School of Physics and Astronomy, University of Leeds, Leeds LS2 9JT, United Kingdom}
\affiliation{$^{2}$Institute of Mathematical Sciences, CIT Campus, Chennai 600113, India}
\affiliation{$^{3}$Homi Bhabha National Institute, Training School Complex, Anushaktinagar, Mumbai 400094, India} 

\date{\today}

\begin{abstract}
The understanding of particle entanglement is an important goal in the studies of correlated quantum matter. The widely-used method of scanning tunneling spectroscopy -- which measures the local density of states (LDOS) of a many-body system by injecting or removing an electron from it -- is expected to be sensitive to particle entanglement. In this paper, we systematically investigate the relation between the particle entanglement spectrum (PES) and the LDOS of fractional quantum Hall (FQH) states, the paradigmatic strongly-correlated phases of electrons with topological order. Using exact diagonalization, we show that the counting of levels in both the LDOS and PES in the Jain sequence of FQH states can be predicted from the composite fermion theory. We point out the differences between LDOS and PES characterization of the bulk quasihole excitations, and we discuss the conditions under which the LDOS counting can be mapped to that of PES. Our results affirm that tunneling spectroscopy is a sensitive tool for identifying the nature of FQH states.
\end{abstract}

\maketitle

\section{Introduction} 

Since their discovery, strongly-correlated topological phases of electrons, such as those in the regime of the fractional quantum Hall (FQH) effect, have continued to attract attention for their exotic properties~\cite{Tsui82, Laughlin83, Halperin82, Jain89, WenEdge, Moore91}, including the recent experimental measurements of fractional statistics of their underlying charged quasiparticles~\cite{Nakamura2020, Bartolomei20}. A promising approach for directly observing these quasiparticles is the scanning tunneling microscopy (STM)~\cite{Papic18}. The STM reveals the local density of states (LDOS) spectrum, which is sensitive to the topological order of the underlying FQH phase. Moreover, the LDOS spectrum changes if the fundamental quasiparticle gets trapped by an impurity. Recent work~\cite{Xiaomeng} successfully employed STM measurements to visualize atomic-scale electronic wave functions, revealing microscopic signatures of valley ordering and spectral features of FQH phases in graphene. In light of these experimental developments, it is important to formulate a general framework for the theoretical interpretation of the LDOS spectra for different classes of FQH phases, in particular going beyond the low-density limit where analytic treatments are possible~\cite{MacDonald2010, Lim2011}.

A highly-accurate approach for describing a large class of FQH states and their low-energy excitations is the composite fermion (CF) theory~\cite{Jain89}. This theory applies primarily to the so-called Jain sequence of states at the electron filling factors $\nu=n/(2pn\pm 1)$, where $n$ and $p$ are positive integers, that in particular capture the most prominent incompressible FQH plateaus observed in the lowest Landau level (LLL). The main tenet of CF theory is that the Jain states can be viewed as integer quantum Hall (IQH) states of CFs~\cite{Jain07}. The latter are bound states of an electrons and an even number ($2p$) of quantized vortices. Importantly, the CF theory not only accurately captures the FQH ground states, but also the entire low-energy spectrum, including both charged as well as charge-neutral excitations (for a recent overview of CF theory, see~\cite{HalperinJainBook}). Given that the LDOS spectrum is determined by the systems' bulk excitations, it is natural to expect that its structure can be predicted and understood, at least qualitatively, from CF theory, as first proposed in Ref.~\cite{Papic18}. In this work, we systematically investigate LDOS for the Jain states through a combination of exact diagonalization simulations and CF theory.

The second motivation behind this paper is to relate the structure of the LDOS spectra with the entanglement of the underlying electrons forming the FQH fluid. The relevant entanglement measure in this context is the so-called \emph{particle entanglement spectrum} (PES), first introduced in Refs.~\cite{Sterdyniak11,Sterdyniak2012} as a generalization of the particle entanglement entropy~\cite{Zozulya2008,Haque09}. To evaluate PES, one performs the Schmidt decomposition on a state by dividing it into two parts, each with a fixed number of particles while the total area and geometry of the system remain unchanged. Recently, PES has been fruitfully studied in rotating two-dimensional gases~\cite{Heng2010}, Bose-Einstein condensates~\cite{Liu10} and Luttinger liquids~\cite{Herdman2015}, Hubbard models~\cite{Herdman14, Iemini15, Ferreira22}, magic-angle twisted bilayer graphene~\cite{Repellin20}, and various lattice models~\cite{Rammelmuller17,Barghathi17} including driven optical lattices~\cite{Hudomal19}. Since PES is obtained by tracing out some particles from the system, it can be intuitively thought of as introducing quasihole excitations to the bulk of the FQH fluid~\cite{Sterdyniak11}. In this sense, PES is conceptually similar to LDOS, however, the relation between the two has not been scrutinized thus far.

The remainder of this paper is organized as follows. In Sec.~\ref{CF}, we give a brief overview of CF theory and explain how it describes the low-energy excitations when the magnetic flux is slightly increased or reduced relative to its value at the center of an FQH plateau, corresponding respectively to creating `quasihole' and `quasiparticle' excitations on top of the ground state. In Sec.~\ref{LDOS}, we introduce LDOS and present the results of its numerical simulations in realistic electron systems with screened Coulomb interaction and in the presence of a charged impurity. We identify the distinctive counting of LDOS, resolved by the angular momentum quantum number in a reference frame centered at the impurity, and we show how this counting can be understood from CF theory. In Sec.~\ref{PES}, we introduce PES and explain how its counting can be derived from CF theory. In Sec.~\ref{discussion}, we arrive at our central question, i.e., the relation between PES and LDOS countings. In particular, we explain the difference between these two, albeit both measure the bulk quasihole excitations, and we discuss the conditions under which the LDOS counting can be mapped to the PES counting (and vice versa). Our results are summarized in Sec.~\ref{conclusion}. In Appendix~\ref{parameters}, we discuss the optimization of the parameters of the screened Coulomb interaction that yield the best conditions for observing the key features of the FQH fluid in LDOS. 

\section{Low-energy excitations from composite fermion theory}
\label{CF}

A broad class of FQH states occurring at filling factors $\nu=n/(2pn\pm 1)$, can be understood as IQH states of CFs filling $\nu^{*}=n$ effective LLs~\cite{Jain89} (a star superscript is used to denote CF quantities). Here we use the spherical geometry~\cite{Haldane83} to briefly illustrate the mapping between FQH states of electrons and IQH states of CFs. In the spherical geometry, the two-dimensional gas of $N$ electrons is placed on the surface of a sphere, with a Dirac monopole of strength $2Q$ flux quanta situated at the center ($2Q$ is a positive integer). The radius of the sphere is $R=\sqrt{Q}$ in units of the magnetic length $\ell=\sqrt{\hbar/eB}$, where $B$ is the strength of the magnetic field pointing radially outwards. On the sphere, the LL indexed by $n$ has a degeneracy of $2(Q+[n-1])+1$ (in our convention, the LLL has index $n=1$.). 

\begin{figure}[tb]
	\includegraphics[width=0.96\columnwidth]{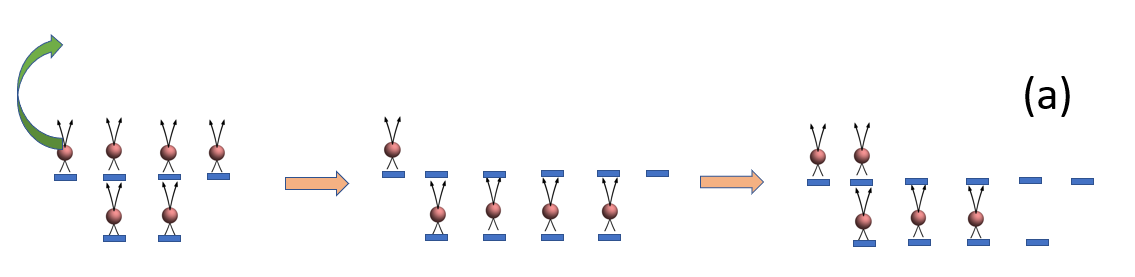}
\includegraphics[width=0.96\columnwidth]{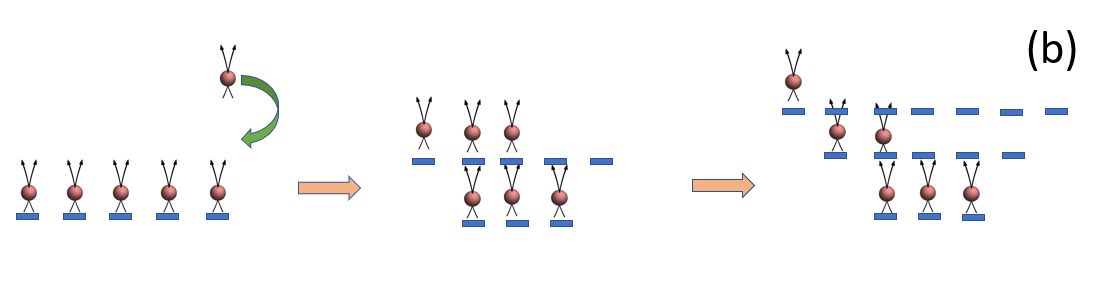} 
\includegraphics[width=0.96\columnwidth]{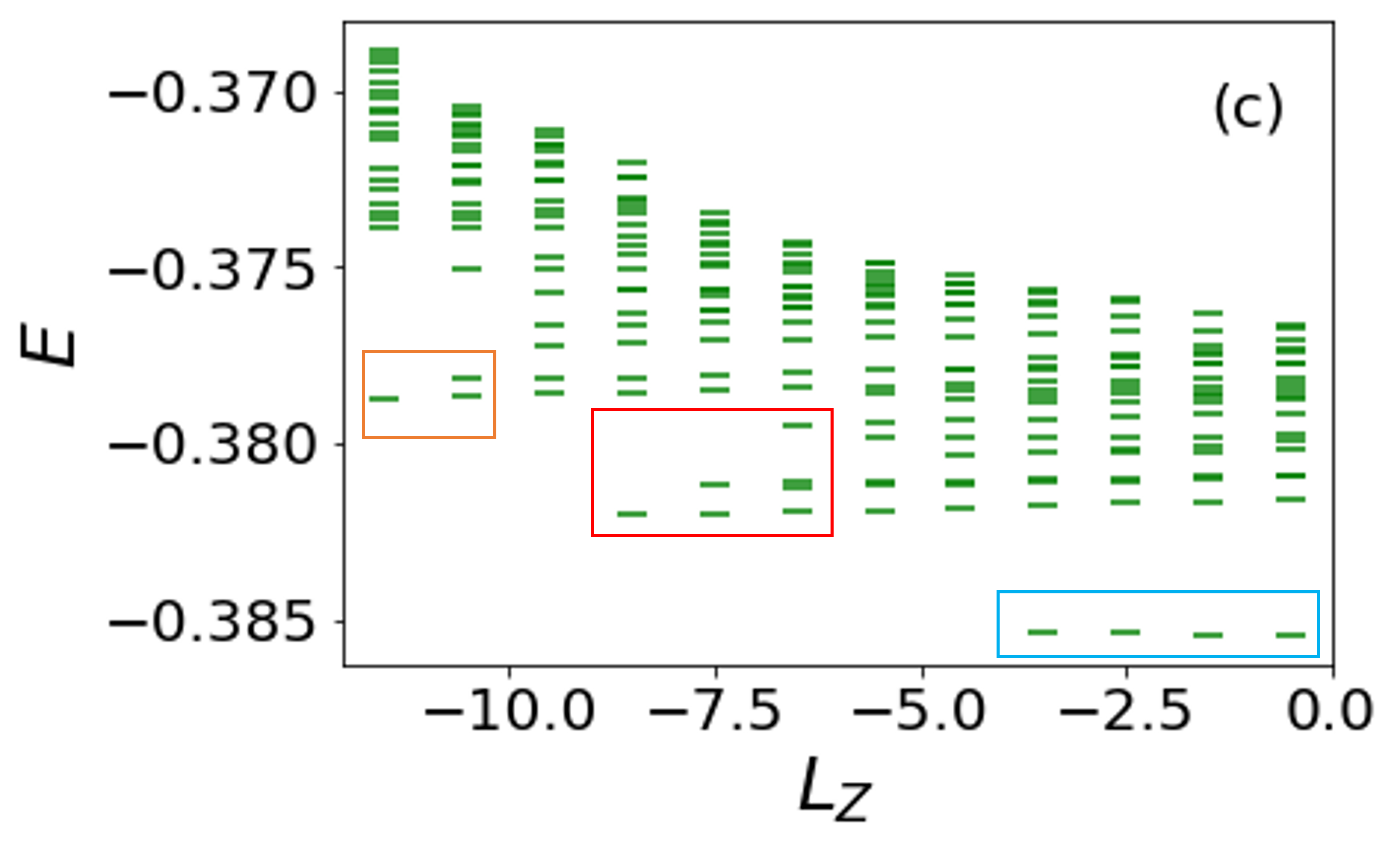}
	\caption{A schematic illustration of removing an electron from $\nu{=}2/5$ (a) and adding an electron to $\nu{=}1/3$ (b). The left column represents the ground states before removing or adding an electron. The middle column shows the starting $m$ state in the lowest branch and the right column shows the starting $m$ state in the second-lowest branch. (c) The energy spectrum for $N=11$ electrons at $2Q=23$ flux quanta in the LLL (computed by exact diagonalization in the spherical geometry). The parameters of the screened Coulomb interaction and impurity potential are given in Eqs.~\eqref{interaction}-\eqref{potential} with $d_g=3$, $ d_i=2.9$. The blue rectangle labels the lowest branch, the red rectangle labels the second branch, and the orange rectangle labels the third branch. This spectrum corresponds to a hole excitation of the $\nu=3/7$ state. As seen in the plot, the lowest branch and part of the second branch and the third branch are separated from other parts by a gap. Energies are quoted in $e^2/\epsilon\ell$ units. 
	}
	\label{fig2}
\end{figure}

In CF theory, every electron binds $2p$ quantized vortices to form a CF. Thereby, the number of effective flux quanta felt by a CF is $2Q^*=2Q-2p(N-1)$. In the zeroth-order approximation, the CFs are taken to be non-interacting particles, and thus in an effective magnetic field, they form Landau-like levels termed ``Lambda levels" ($\Lambda$Ls). The formation of CFs captures the leading effect of the electron-electron interactions and lifts the degeneracy of the lowest LL. Since $|Q^*|<Q$, each $\Lambda$L accommodates fewer states than the electronic LL. If an integer number of $\Lambda$Ls are filled, a gapped ground state can be formed. In this case, the flux-particle relationship at $\nu=n/(2pn\pm 1)$ is
\be
N=\frac{n}{2pn \pm 1}\left(2Q\pm n+2p\right).
\ee
The $+$ ($-$) sign in $\nu=n/(2pn\pm 1)$ denotes that the effective magnetic field sensed by the CFs is in the same (opposite) direction as the external magnetic field seen by the electrons.

The FQH ground state at $\nu=n/(2pn\pm 1)$ is represented by the Jain wave function
\be
\Psi^{\rm Jain}_{\frac{n}{2pn \pm 1}}=\mathcal{P}_{\rm LLL}\Phi_{\pm n}\Phi_1^{2p},
\label{eq: Jain_wf}
\ee
where $\Phi_{n}$ is the $n$-filled LL wave function ($\Phi_{-n} \equiv \Phi_n^*$). For $n=1$ and the $+$ sign in Eq.~\eqref{eq: Jain_wf}, the Jain wave function reduces to the Laughlin wave function $\Psi^{\rm Laughlin}_{1/(2p+1)}=\Phi_1^{2p+1}$~\cite{Laughlin83}. Except for the Laughlin case, the product state $\Psi_{\pm n}\Psi_1^{2p}$ does not reside fully in the LLL, so a projection operator $\mathcal{P}_{\rm LLL}$ onto the LLL is needed to describe the state that arises in the high-field limit~\cite{Girvin84b}. 

The low-energy excitations can be obtained by replacing $\Phi_{n}$ with low-energy excitations of $n$-filled LLs. This provides an efficient way to count the low-energy excitations in different branches, which are separated by the effective cyclotron energy of the $\Lambda$Ls, as illustrated in Fig.~\ref{fig2}. Panels (a) and (b) of Fig.~\ref{fig2} show the $\Lambda$L occupation for CF hole and CF particle excitations of different branches, and panel (c) shows an example of the energy spectrum where the lowest branch can be fully identified while the second and third branches can be partially identified. 

Nevertheless, when the CF interaction plays a nonperturbative role, the na\"ive counting based on non-interacting CFs can be misleading. This can be intuitively understood since the na\"ive counting of non-interacting CFs would predict an infinite number of excitations for a finite system in the LLL (since one can create excitations in arbitrarily high $\Lambda$Ls), which cannot be true. The CF-CF interactions make certain excitations prohibitively expensive to create, so they should be excluded from the counting. A comprehensive discussion of the exclusion rules that arise from the residual interaction between CFs is presented in Ref.~\cite{Balram13}. As shown in Ref.~\cite{Balram13}, for bare quasiparticle and quasihole excitations that are not dressed by neutral excitations, the na\"ive IQH counting (without the need to impose any exclusion rules) works, which enables us to develop the LDOS counting conveniently for these excitations.

We now illustrate the counting for excited states in the CF theory. We first consider Jain states at $\nu=n/(2pn+1)$. When an electron is removed from the $\nu=n/(2pn+1)$ Jain state, the effective magnetic field is increased, i.e., $2Q^*\rightarrow 2Q^*+2p$. In other words, each $\Lambda$L has $2p$ more orbitals. Together, these create $2pn+1$ CF holes (the additional one accounts for the removed electron) in the lowest $n$ $\Lambda$Ls, as shown in Fig.~\ref{fig2}(a). The lowest branch of the excited states will have all the CF holes in the topmost occupied $\Lambda$L with index $n$. Therefore, the counting of the lowest branch is given by placing $2pn+1$ holes in $2[Q^*+(n-1)]+1$ orbitals. The largest value of the magnitude of $L_z$, the $z$-component of the total orbital angular momentum, of these states is $\sum_{l=0}^{2pn}\left(Q^*+(n-1)-l\right)$ (we have set $\hbar=1$), and a representative of the corresponding configuration is illustrated in the middle panel of Fig.~\ref{fig2}(a). On the other hand, for states with one electron removed from the ground state, the largest possible $L_z=Q$ since the ground state has $L_z=0$. The difference between the smallest (and largest) $L_z$ of all excitations and excitations in the lowest branch is 
\ba
\label{startLz-hole}
\nonumber m_0 &\equiv& Q-\sum_{l=0}^{2pn}\left(Q^*+(n-1)-l\right) \\
&=& (2p^2-p)n^2+(3p-2p^2-1)n-2p+1,
\ea
which is a number independent of system size. The excitation with the lowest $L_z$ would be obtained by removing an electron from the origin, which has $L_z=0$ when the sphere is mapped to the infinite plane in the thermodynamic limit. Therefore, the lowest $L_z$ state has the same $L_z$ value as the ground state, and we define this as $m=0$. Here $m$ represents the difference of $L_z$ of the excited state with respect to the ground state in the thermodynamic limit. Clearly, Eq.~\eqref{startLz-hole} gives the starting $m$ for the lowest branch of excitations. As we will show in Sec.~\ref{LDOS}, this quantity is directly accessible in the STM spectrum.

Moreover, if we denote the number of states in the $L_z$ sector with $N_h$ holes or particles in a $\Lambda$L with $M$ orbitals as $(L_z,N_h,M)$, the counting for the lowest branch of excitations is given by 
\begin{eqnarray}
\label{LDOS-hole-1}
\nonumber \left(L_z,2pn+1,\frac{2Q}{2pn+1}+\frac{2pn^2-2pn+4p+2n-1+4p^2n}{2pn+1}\right). \\
\end{eqnarray}
We impose the restriction that $(L_z,N_h,M)=0$ if $|L_z|>(N_h/2)(M-N_h+1)$. In general, $(L_z,N_h,M)$ can be computed numerically once specific values are given. This number is essentially the number of different partitions of $L_z$ into $N_h$ different integers or half-integers, with each integer in the range of $\left[-(M-1)/2, (M-1)/2\right]$ or half-integer in the range of $\left[-M/2, M/2\right]$. 

For the second lowest branch, we put $2pn$ holes in the $n$th $\Lambda$L and one hole in the $(n-1)$th $\Lambda$L. The starting $m$ is decreased by $2pn-1$, as shown in the right column of Fig.~\ref{fig2}(a). The counting in the second lowest branch in the $L_z$ sector is given by 
\begin{widetext}
\be
\sum_{q=-(Q^*+n-1)}^{Q^*+n-1} \left(L_z-q,2pn,\frac{2Q}{2pn+1}+\frac{2pn^2-2pn+4p+2n-1+4p^2n}{2pn+1}\right).
\label{LDOS-hole-2}
\ee
\end{widetext}

Similarly, when an electron is added to the $\nu=n/(2pn+1)$ Jain state, the effective magnetic field is decreased, i.e., $2Q^*\rightarrow 2Q^*-2p$. Along with the newly added electron, there are $2pn+1$ electrons in the $(n+1)$th and higher $\Lambda$Ls, as shown in Fig.~\ref{fig2}(b). The lowest branch will have all the electrons in the $(n+1)$th $\Lambda$L. The starting $m$ is
\ba
\notag m_0 &=& Q-\sum_{l=0}^{2pn}\left(Q^*+n-l\right)\\
&=& (2p^2-p)n^2+(2p^2+p-1)n.
\label{startLz-particle}
\ea
The counting in each $L_z$ sector is given by 
\be
\left(L_z,2pn+1,\frac{2Q}{2pn+1}+\frac{2pn^2+2pn+2n-4p^2n+1}{2pn+1}\right).
\label{LDOS-particle-1}
\ee
Analogously, we obtain the second branch by putting $2pn$ electrons in the $(n+1)$th $\Lambda$L and one electron in the $(n+2)$th $\Lambda$L. The starting $m$ is decreased by $2pn+1$. The counting in each $L_z$ sector is given by
\begin{widetext}
\be
\sum_{q=-(Q^*+n+2)}^{Q^*+n+2} \left(L_z-q,2pn,\frac{2Q}{2pn+1}+\frac{2pn^2+2pn+2n-4p^2n+1}{2pn+1}\right).
\label{LDOS-particle-2}
\ee
\end{widetext}

Having derived the particle and hole counting for $\nu=n/(2pn+1)$ Jain states, it is straightforward to understand the counting for the $\nu=n/(2pn-1)$ Jain states. Adding one electron will increase $2|Q^*|$ by $2p$, thereby creating $2pn-1$ quasiholes, while removing one electron will reduce $2|Q^*|$ by $2p$, thereby creating $2pn-1$ quasiparticles. For the special case of $p=1$, the $\nu=n/(2n-1)$ state is just the particle-hole conjugate of the $\nu=(n-1)/[2(n-1)+1]$ state. Therefore, the particle counting for $\nu=n/(2n-1)$ is identical to the hole counting for $\nu=(n-1)/[2(n-1)+1]$, and vice versa.
For instance, when we remove an electron from the $\nu=2/3$ ground state, $Q^*$ increases by 2. However, as $Q^*$ is a negative number, its magnitude decreases by 2. There are four fewer orbitals in the two occupied $\Lambda$Ls, with an electron removed. This creates three particles in the higher $\Lambda$Ls, which is also the case for adding an electron to $\nu=1/3$. For more general $\nu=n/(2pn-1)$ Jain states, the counting is given by the number of ways of distributing the $2pn-1$ quasiholes or quasiparticles in the $\Lambda$Ls, which can be derived similarly to that for the $\nu=n/(2pn+1)$ states discussed above.

\section{Local density of states}
\label{LDOS}
Recently, it has been proposed that the fractional statistics of excitations in FQH states can be visualized using STM~\cite{Papic18}. The STM directly probes the LDOS of an FQH state by measuring the differential conductance~\cite{He1993, Hatsugai1993, Efros1993, Johansson1993, Varma1994, Aleiner1995, Haussmann1996}. 
The LDOS for removing or adding an electron to the ``vacuum" state $|\Omega\rangle$ is defined as
\be
{\rm LDOS}(E,m)= 
\begin{cases}
\sum_a \delta(E-E_a)|\langle a|c_m|\Omega\rangle|^2,
\;\;\; \mathrm{hole \; side}, \\ 
\sum_a \delta(E-E_a)|\langle a|c_m^\dagger|\Omega\rangle|^2, \;\;\; \mathrm{particle \; side}.
\end{cases}
\ee
Here $|a\rangle$ runs over all energy eigenstates corresponding to the FQH state with one electron removed or injected from $|\Omega\rangle$, and $E_a$ is the corresponding eigenvalue (measured relative to the ground state energy of $|\Omega\rangle$). The operators $c_m$ and $c_m^\dagger$ destroy and create, respectively, an electron in a LLL orbital with angular momentum $L_z =m$. The resolution of LDOS into $m$ sectors allows us to conveniently study its discrete counting. However, this relies on a relatively low concentration of impurities (in our numerics below, we consider only a single impurity), which do not fully destroy the translation or $z$-rotation invariance of the system. More generally, one can compute LDOS in real space by replacing $c_m$, $c_m^\dagger$ with the corresponding field operators and then perform a ``Laguerre transform" to approximately extract the counting per $m$ sector~\cite{Papic18}.

In the numerical simulations below, we evaluate LDOS for a system of electrons in the spherical geometry~\cite{Haldane83, Fano86} using the model introduced in Ref.~\cite{Papic18}. The electrons interact via screened Coulomb potential given by
\be
\label{interaction}
V_C(r)=\frac{1}{r}-\frac{1}{\sqrt{(2d_g)^2+r^2}},
\ee
where the screening distance $d_g$ represents the distance between the FQH system and a metallic gate. 
The one-body potential due to an impurity at a distance $d_i$ from the electron gas ($d_i<d_g$) is given by
\be
\label{potential}
U(r)=\frac{Z}{\sqrt{d_i^2+r^2}}-\frac{Z}{\sqrt{(2d_g-d_i)^2+r^2}},
\ee
where $Z{=}1$ ($-1$) for repulsive (attractive) impurity potential placed at the South (North) Pole which can bind quasiholes (quasiparticles). Below we quote all lengths in units of the magnetic length $\ell$ and the energies are quoted in units of $e^2/\epsilon\ell$. We assume the magnetic field is strong enough such that the kinetic energy is quenched and the electrons are confined in the LLL.

In our calculations, $d_i$ and $d_g$ are treated as tunable parameters. As shown in Appendix~\ref{parameters}, we can adjust the values of $d_g$ and $d_i$ to separate the different branches in LDOS and facilitate their identification. We find the optimal parameters to be $d_g\sim 3\ell$ and $d_i\sim 2.9 \ell$ for all filling factors considered in our study. We also note that the weight of LDOS levels is generally not uniformly distributed. The higher-energy LDOS states can have more weight than the lower-energy states. In an STM experiment, a level must have a sufficiently large weight for it to be observable. In our calculations, we keep the weight of all identifiable levels to be no smaller than $10^{-7}$ but do not further consider its distribution in finding the optimal parameters. In all LDOS figures, the background is manually set to $\sim 10^{-12}$ to provide a large contrast in the color to assist in identifying the counting. 

As explained in the previous section, when an electron is injected or removed from the bulk of a Jain state at $\nu=n/(2pn\pm 1)$, there are $(2pn\pm 1)$ anyons (quasiparticles or quasiholes) created. If there is a single isolated impurity, the LDOS spectrum directly shows the discrete energy levels of anyons bound to the impurity. The spectrum is determined by the fractional exclusion statistics of the anyons~\cite{Haldane1991}, and it is a manifestation of the topological nature of FQH states. Therefore, STM can be a powerful experimental approach to effectively ``image" anyons. 

\begin{figure*}
\includegraphics[width=0.38\textwidth]{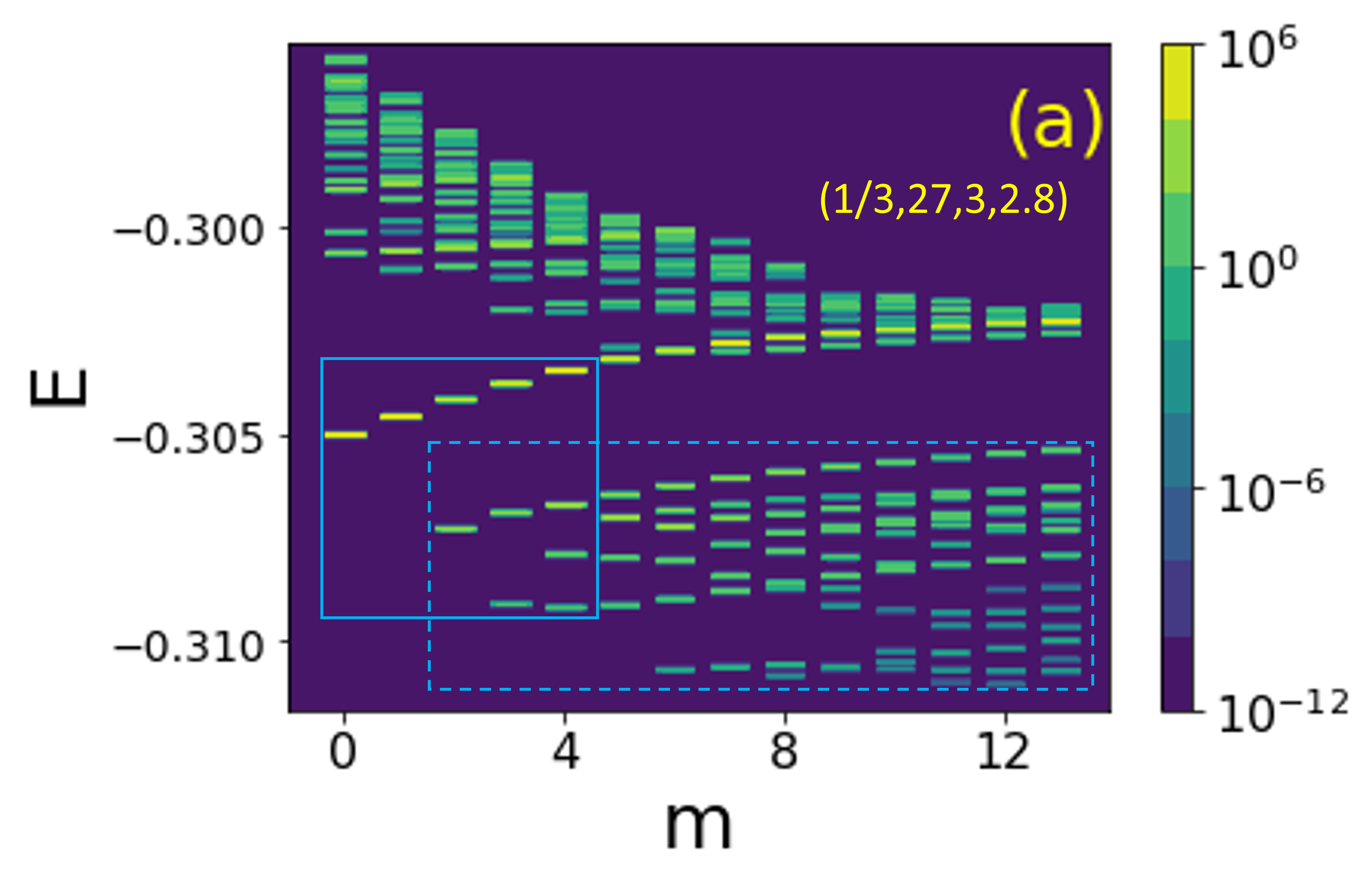}
\includegraphics[width=0.38\textwidth]{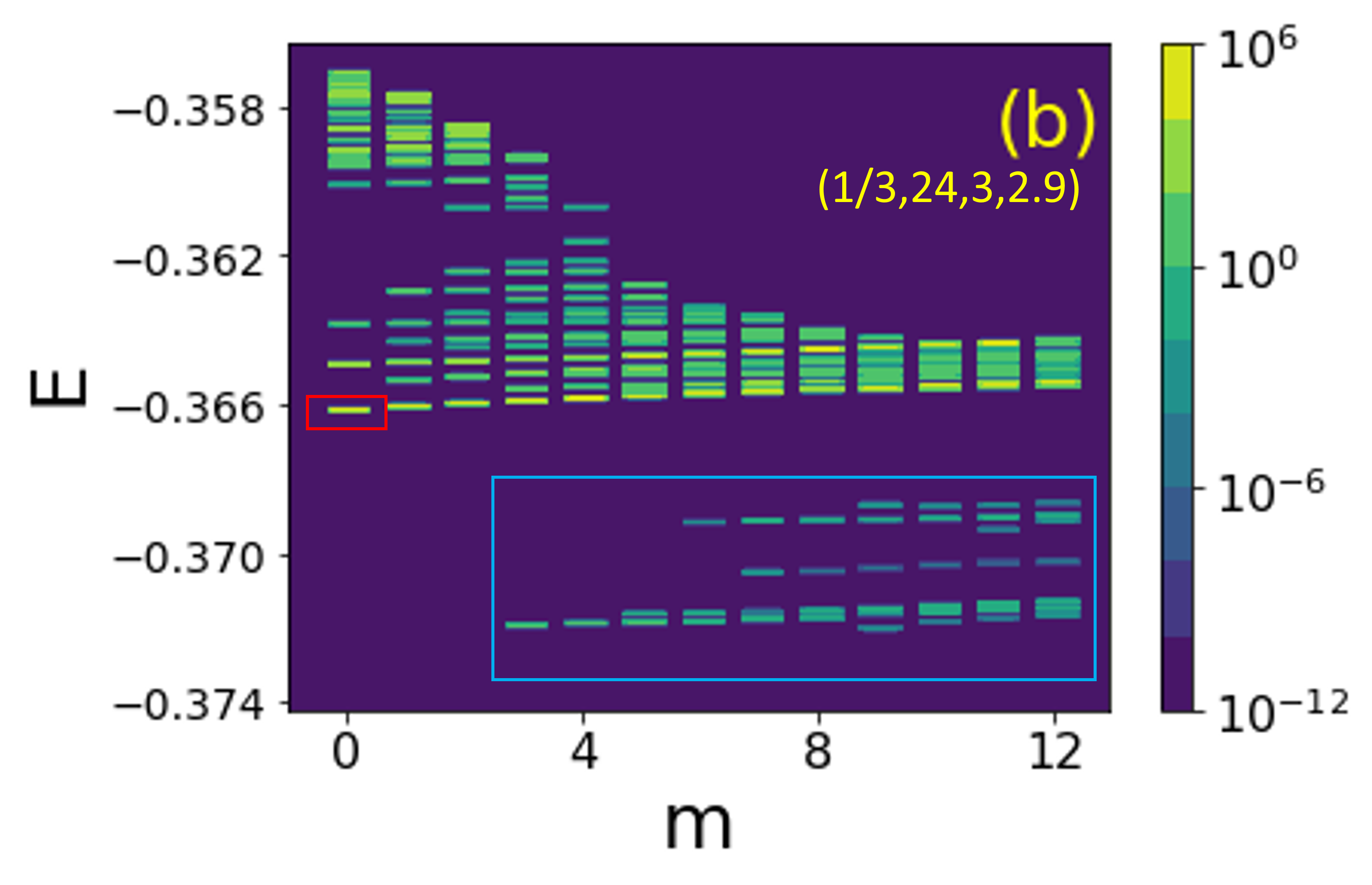}
\includegraphics[width=0.38\textwidth]{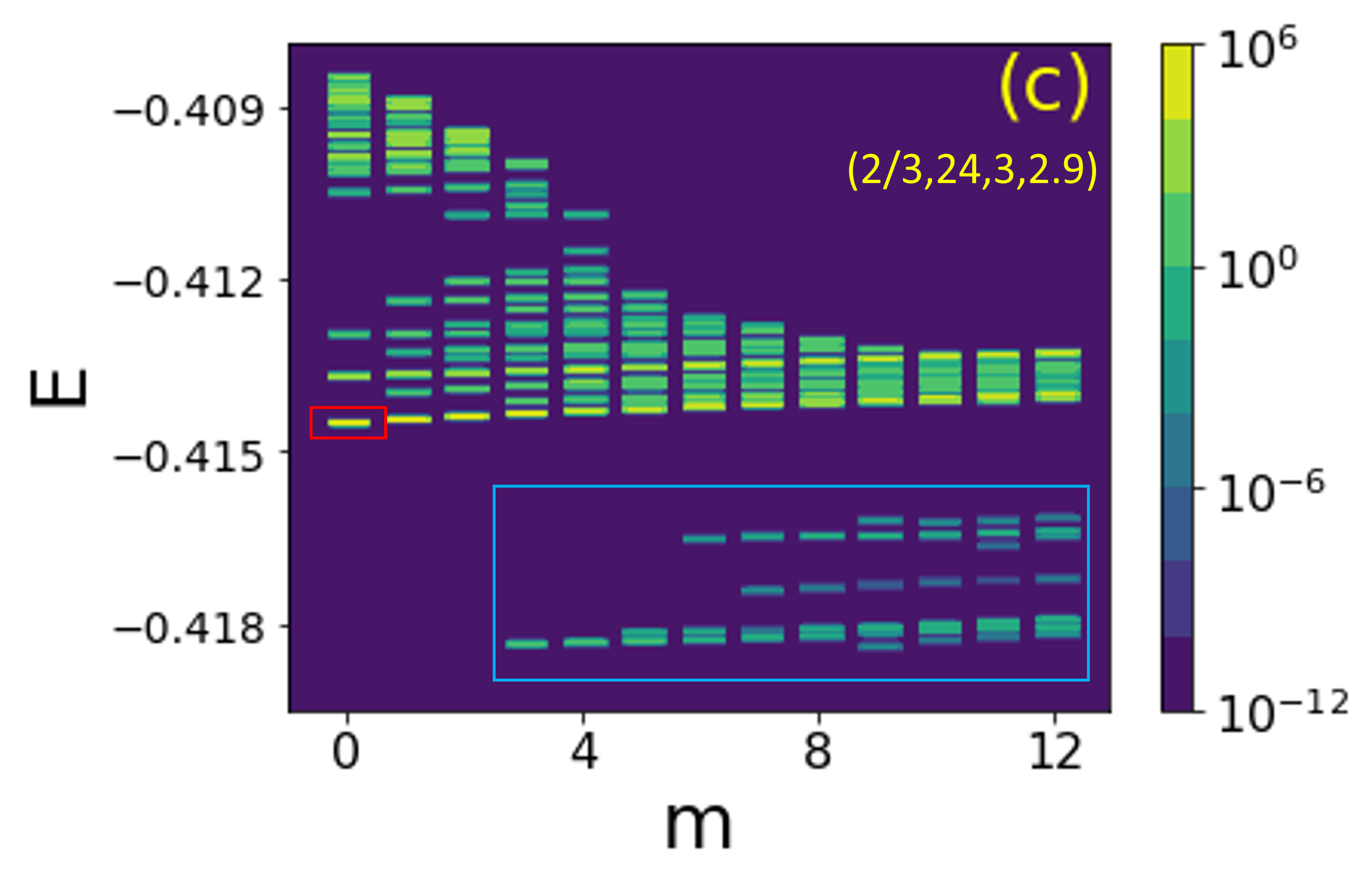}
\includegraphics[width=0.38\textwidth]{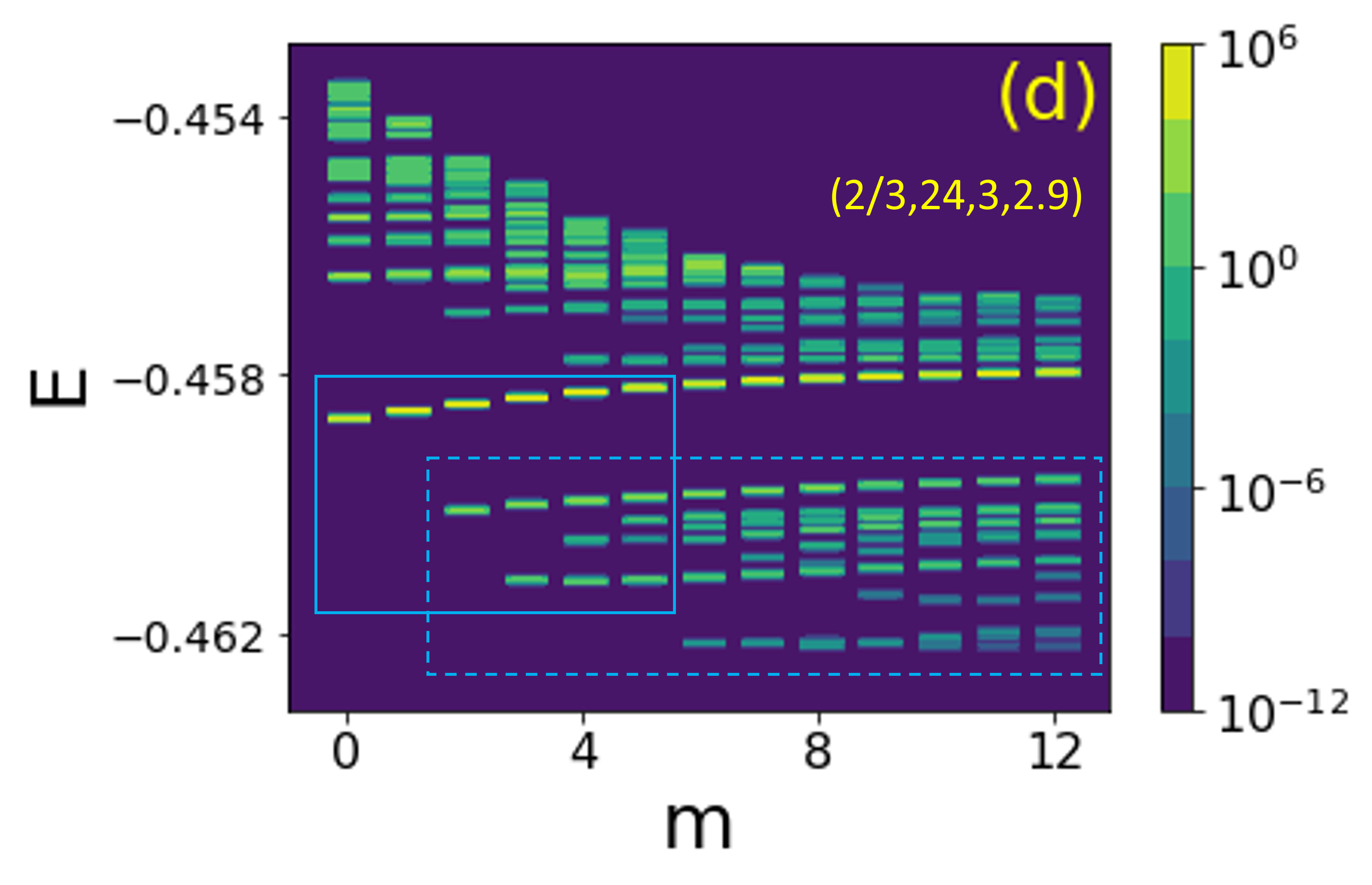}
\includegraphics[width=0.38\textwidth]{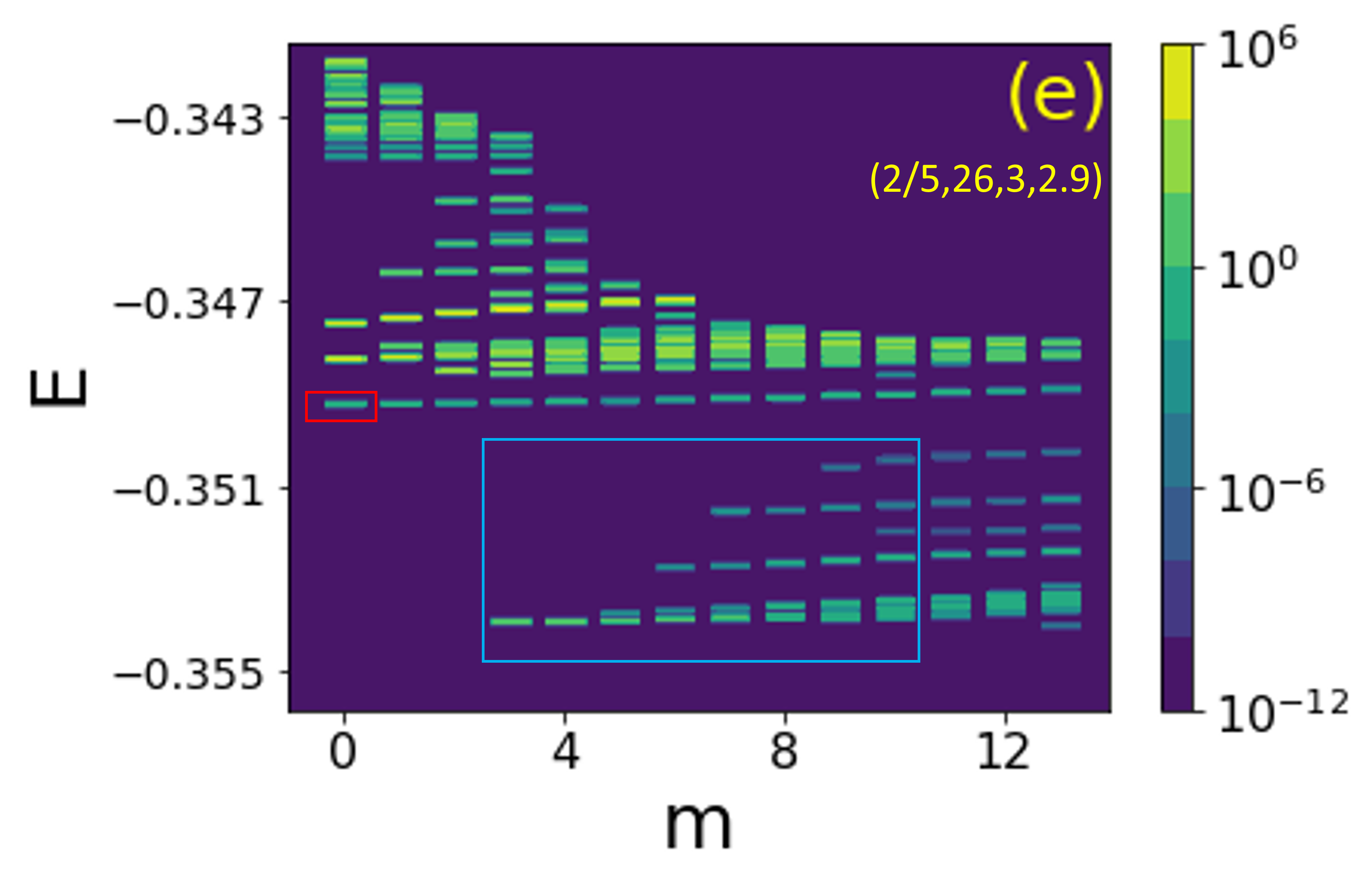}
\includegraphics[width=0.38\textwidth]{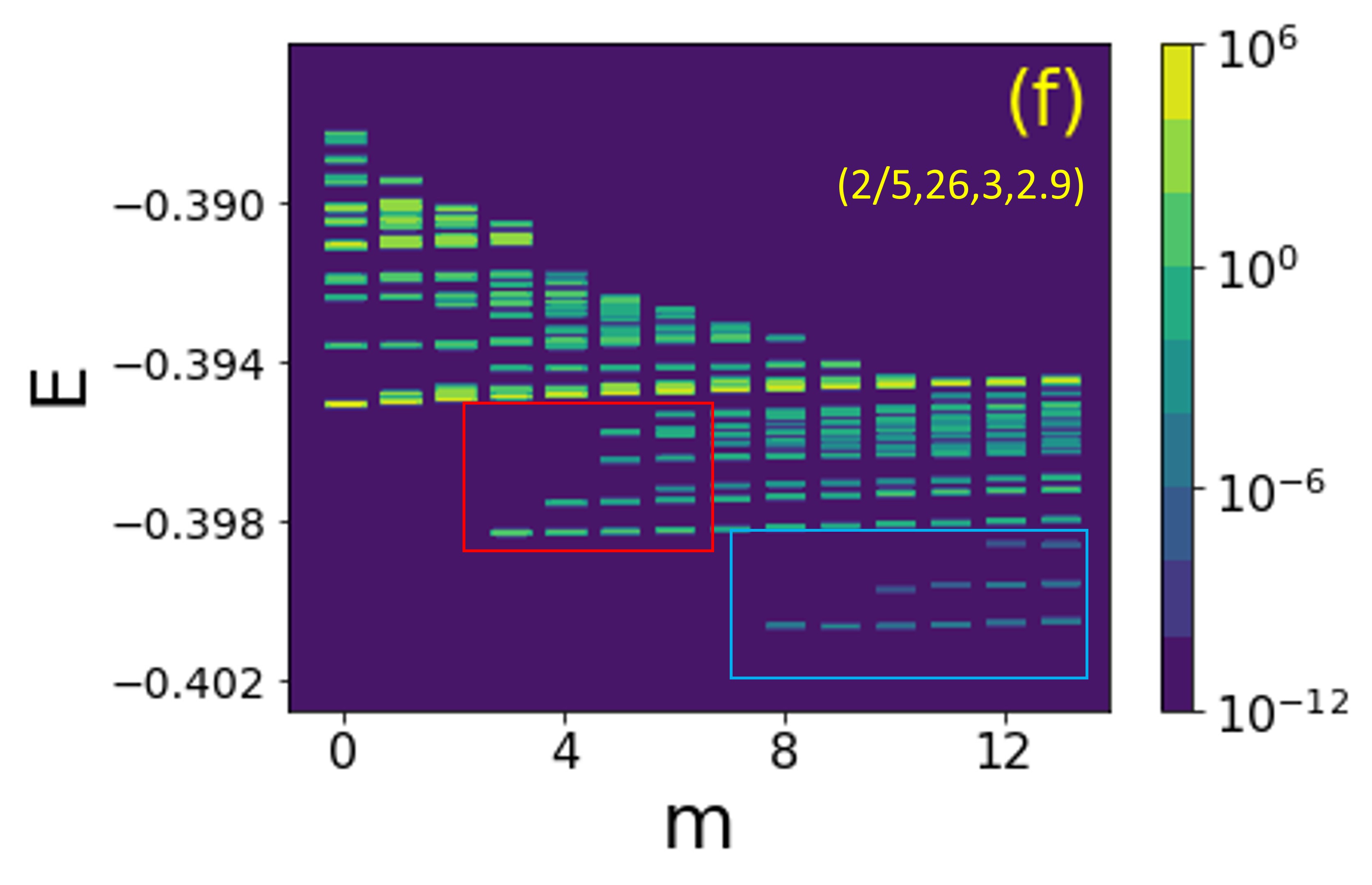}
\includegraphics[width=0.38\textwidth]{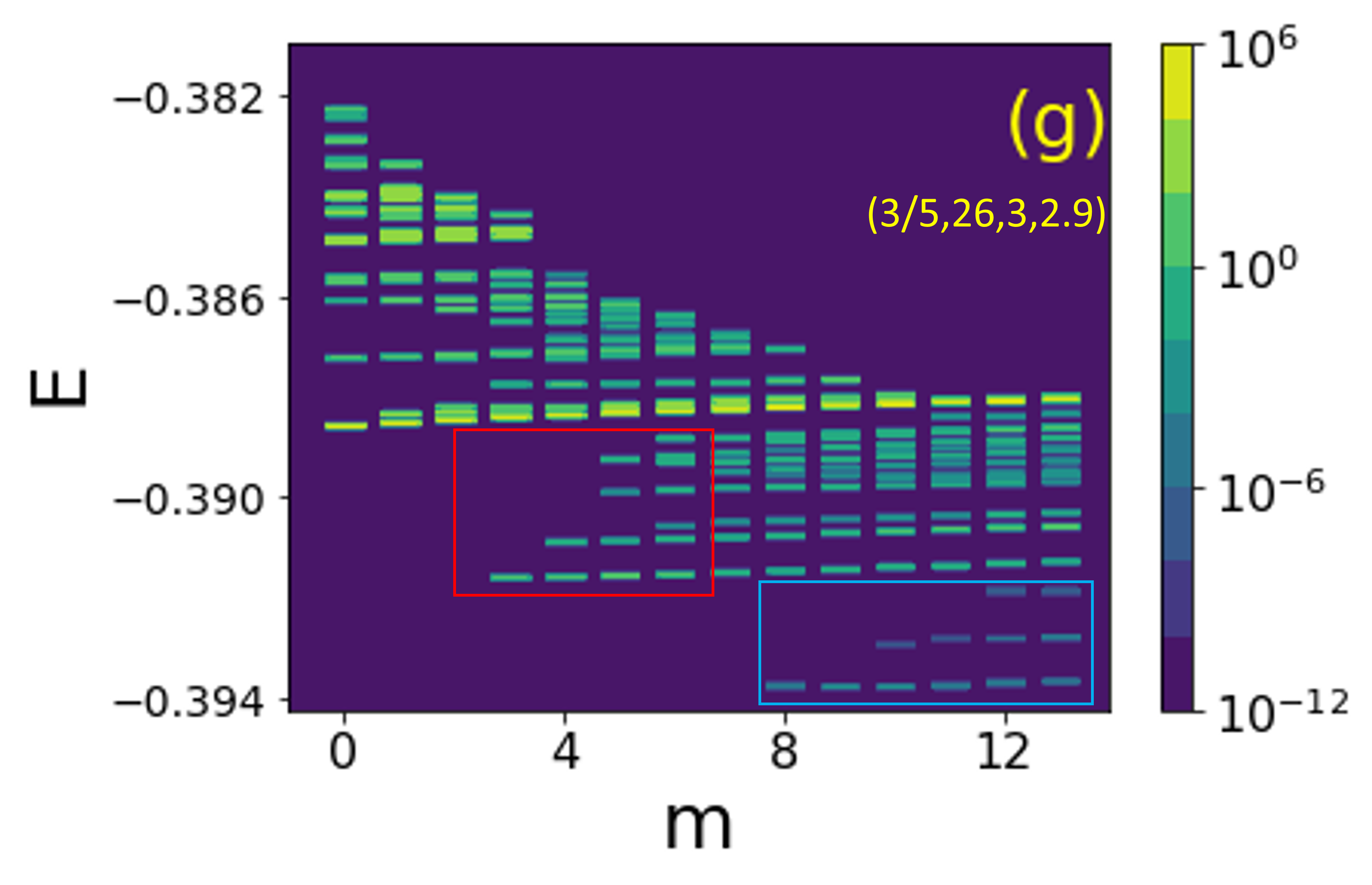}
\includegraphics[width=0.38\textwidth]{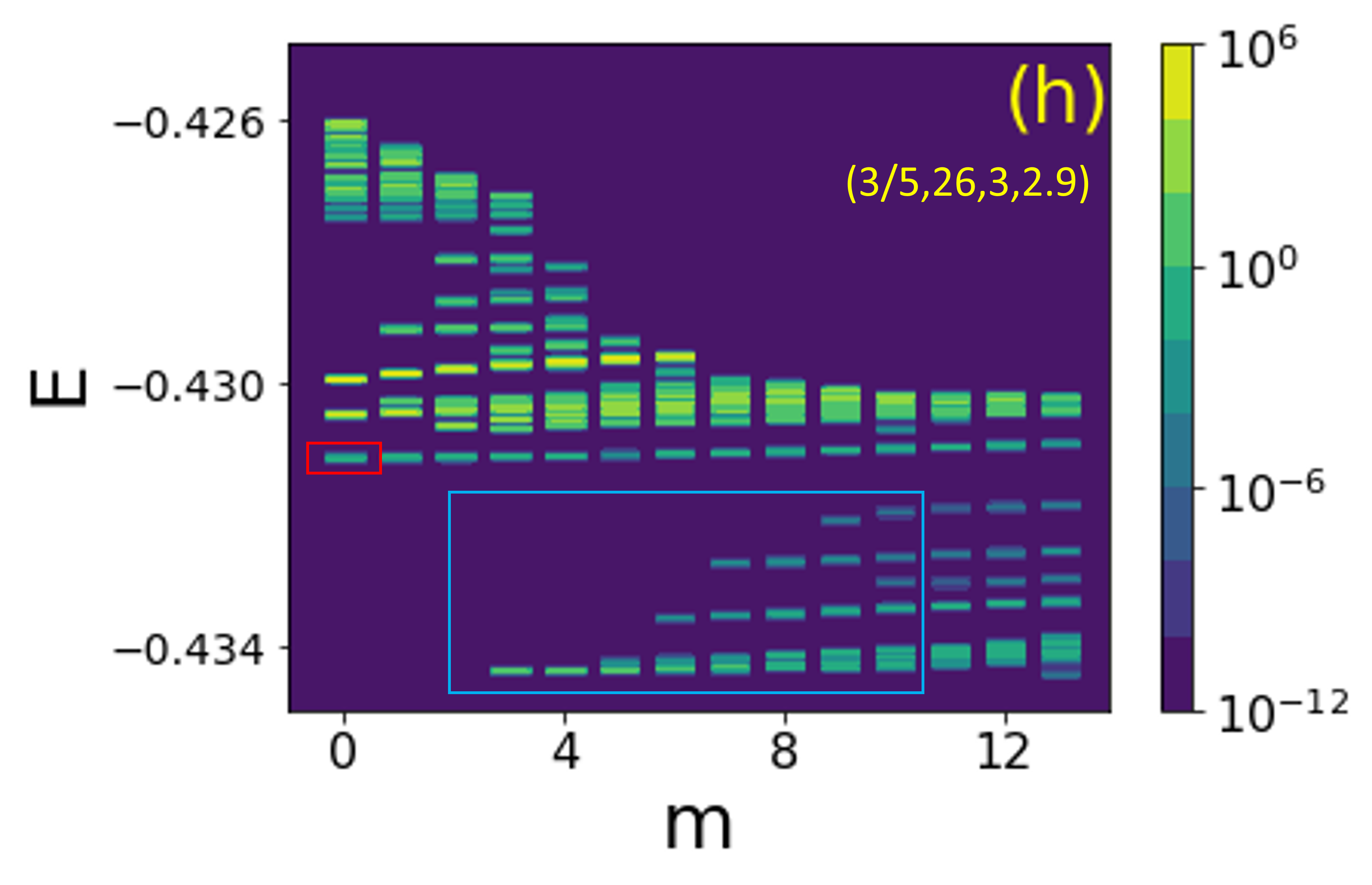}
\includegraphics[width=0.38\textwidth]{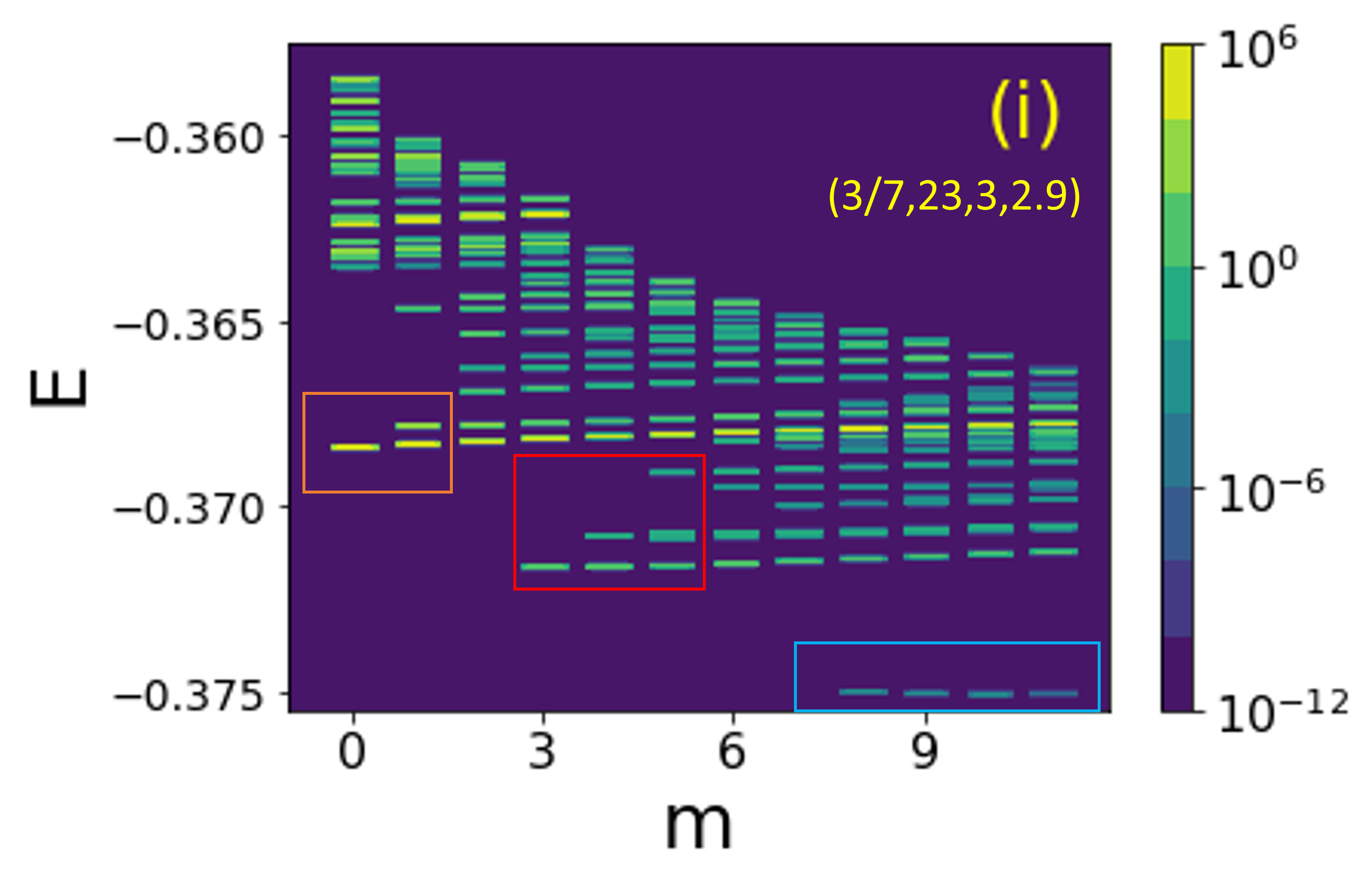}
\includegraphics[width=0.38\textwidth]{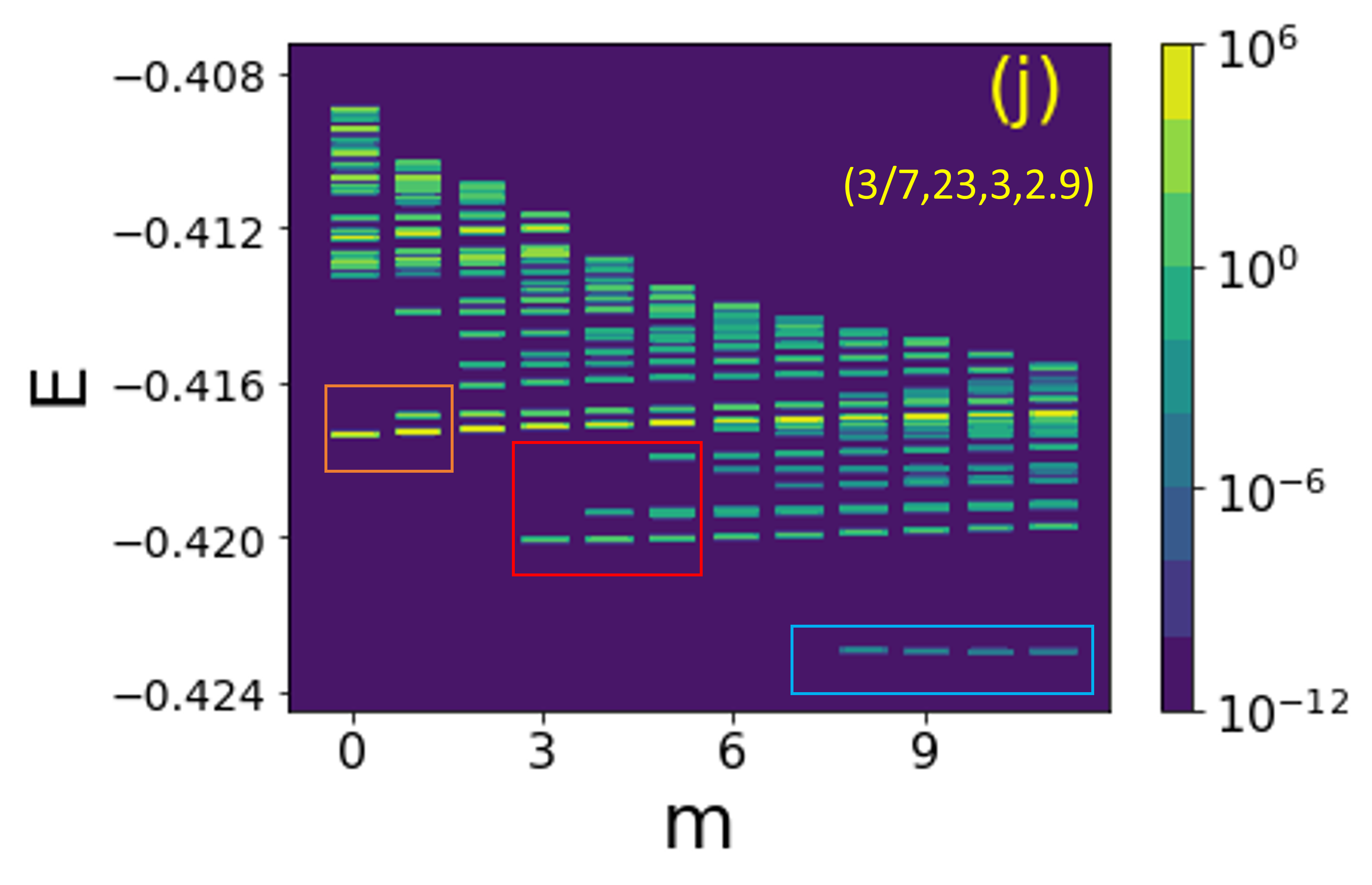}
	\caption{The LDOS spectrum for $\nu=1/3$, $2/3$, $2/5$, $3/5$ and $3/7$ quasihole excitations (left column) and quasiparticle excitations (right column) using the screened Coulomb interaction in Eq.~(\ref{interaction}) and impurity potential in Eq.~(\ref{potential}). Plots are labelled according to $(\nu, 2Q, d_g, d_i)$. The lowest three branches are denoted by blue, red, and orange rectangles. In panels (a) and (d), the dashed blue rectangle denotes the lowest branch \emph{without} its highest-energy multiplet.}
	\label{LDOS2_5}
\end{figure*}

To interpret the LDOS spectrum, we appeal to the CF theory which, as explained above, accurately describes the low-energy excitations of Jain states. Therefore, this theory also makes predictions for the counting in LDOS spectra. For the low energy part of the LDOS spectrum, the starting $m$ is given by Eq.~\eqref{startLz-hole} and Eq.~\eqref{startLz-particle}, and the counting in each $L_z$ sector is given by Eq.~\eqref{LDOS-hole-1}, Eq.~\eqref{LDOS-hole-2}, Eq.~\eqref{LDOS-particle-1}, and Eq.~\eqref{LDOS-particle-2}. 
In the remainder of this section, we numerically compute the LDOS spectrum using exact diagonalization and compare it to the predictions of CF theory. The LDOS spectrum of the $\nu{=}1/3$, $2/3$, $2/5$, $3/5$, and $3/7$ state in the LLL is shown in Fig.~\ref{LDOS2_5}, and the corresponding countings are summarized in Table.~\ref{table1}.

\begin{table*}
\makebox[\textwidth]{\begin{tabular}{|c|c|c|c|c|}
\hline
$\nu$&type&$2Q$&1st branch&2nd branch\\ \hline
\multirow{4}{*}{$1/3$}&\multirow{2}{*}{h}&\multirow{2}{*}{27}&1,1,2,3,4...&N/A\\\cline{4-5}
&&&(1,1,2,3,4,5,7,8,10,12,13,14,15,15)&N/A\\\cline{2-5} 
&\multirow{2}{*}{p}&\multirow{2}{*}{24}&0,0,0,1,1,2,3,4,5,7,7,8,8&1...\\\cline{4-5}
&&&(0,0,0,1,1,2,3,4,5,7,7,8,8)&(1...)\\\hline 
\multirow{4}{*}{$2/3$}&\multirow{2}{*}{h}&\multirow{2}{*}{24}&0,0,0,1,1,2,3,4,5,7,7,8,8&1...\\\cline{4-5}
&&&(0,0,0,1,1,2,3,4,5,7,7,8,8)&(1...)\\\cline{2-5} 
&\multirow{2}{*}{p}&\multirow{2}{*}{24}&1,1,2,3,4,5...&N/A\\\cline{4-5}
&&&(1,1,2,3,4,5,7,8,10,12,13,14,15,15)&N/A\\\hline
\multirow{4}{*}{$2/5$}&\multirow{2}{*}{h}&\multirow{2}{*}{26}&0,0,0,1,1,2,3,5,6,8,9...&1...\\\cline{4-5}
&&&(0,0,0,1,1,2,3,5,6,8,9...)&(1...)\\\cline{2-5} 
&\multirow{2}{*}{p}&\multirow{2}{*}{26}&0,0,0,0,0,0,0,0,1,1,2,2,3,3&0,0,0,1,2,4,7...\\ \cline{4-5}
&&&(0,0,0,0,0,0,0,0,1,1,2,2,3,3)&(0,0,0,1,2,4,7...)\\\hline
\multirow{4}{*}{$3/5$}&\multirow{2}{*}{h}&\multirow{2}{*}{26}&0,0,0,0,0,0,0,0,1,1,2,2,3,3&0,0,0,1,2,4,7...\\\cline{4-5}
&&&(0,0,0,0,0,0,0,0,1,1,2,2,3,3)&(0,0,0,1,2,4,7...)\\\cline{2-5} 
&\multirow{2}{*}{p}&\multirow{2}{*}{26}&0,0,0,1,1,2,3,5,6,8,9...&1...\\ \cline{4-5}
&&&(0,0,0,1,1,2,3,5,6,8,9...)&(1...)\\\hline
\multirow{4}{*}{$3/7$}&\multirow{2}{*}{h}&\multirow{2}{*}{23}&0,0,0,0,0,0,0,0,1,1,1,1&0,0,0,0,1,2,4...\\\cline{4-5}
&&&(0,0,0,0,0,0,0,0,1,1,1,1)&(0,0,0,1,2,4...)\\\cline{2-5} 
&\multirow{2}{*}{p}&\multirow{2}{*}{23}&0,0,0,0,0,0,0,0,1,1,1,1&0,0,0,0,1,2,4...\\ \cline{4-5}
&&&(0,0,0,0,0,0,0,0,1,1,1,1)&(0,0,0,1,2,4...)\\\hline
\end{tabular}}
\caption{The summary of the lowest branch counting and second branch counting for the interaction parameters used in Fig.~\ref{LDOS2_5}. The numbers in the brackets are predicted by CF theory, while other numbers are from the computed LDOS spectra in Fig.~\ref{LDOS2_5} and Sec.~\ref{LDOS}. The numbers which do not match or are not identifiable due to the absence of gaps have been omitted. \label{table1}}
\end{table*}

At filling $\nu=1/3$, we first confirmed the number of LDOS levels of the exact Laughlin state fully matches the number of CF hole excitations. The LDOS of the exact Laughlin state is obtained from its parent Hamiltonian -- the $V_1$ Haldane pseudopotential~\cite{Haldane83, Prange87}. The complete agreement between the exact LDOS counting and CF theory, in this case, is expected since both the Laughlin state and its hole excitations are zero-energy eigenstates of the $V_1$ interaction. 

Next, in Fig.~\ref{LDOS2_5} (a) and (b) we show the LDOS obtained for the screened Coulomb interaction of Eq.~(\ref{interaction}) at $\nu{=}1/3$. We see the lowest branch of quasiparticle excitations in Fig.~\ref{LDOS2_5}(b) matches the CF theory prediction. For quasihole excitations in Fig.~\ref{LDOS2_5}(a), we see there is a gap that allows us to identify the lowest branch counting with CF theory up to $m{=}5$. For $m{>}5$, the branch merges with the continuum and can no longer be visually identified. One interesting observation is that there is a clear gap separating the highest energy state in the lowest branch from the rest of the lowest branch for all $m$ sectors, as marked by the blue dashed rectangle. In other words, it is \emph{only the highest energy multiplet} in the lowest branch that merges with the continuum of the spectrum for large-$m$ sectors. This gap comes from the interactions between CFs, which lift the degeneracy in the lowest $\Lambda$L. The presence of this gap implies that the highest-energy multiplet in the $L^2=Q^*(Q^*+1)$ sector in the lowest $\Lambda$L is separated from the other states in the lowest $\Lambda$L and merges with the higher excitation. It is unclear why this happens for our specific choice of interactions. 

Figs.~\ref{LDOS2_5}(c) and (d) show the LDOS spectrum for $\nu{=}2/3$, which is the particle-hole conjugate of the $\nu{=}1/3$ state. Hence, the counting for the quasihole excitations matches perfectly with CF theory, while the quasiparticle excitations match up to $m{=}5$. Once again, in Fig.~\ref{LDOS2_5}(d) we see a clear gap that separates the highest-energy state in the lowest branch from the rest. If we exclude this highest-energy multiplet, the counting matches perfectly with CF theory for all $m$ sectors. 

In the case of $\nu{=}1/3$ quasihole excitations and $\nu{=}2/3$ quasiparticle excitations, only one $\Lambda$L is involved and there is only one branch of excitations in CF theory. For the $\nu{=}1/3$ quasiparticle excitations and $\nu{=}2/3$ quasihole excitations, higher branches of excitations are expected. From the LDOS spectrum, we can only see the starting $m$ in the second branch is zero, which matches CF theory. However, we cannot identify further countings because there are no clear gaps. 

Fig.~\ref{LDOS2_5}(e) shows that the counting of the lowest branch of LDOS for $\nu{=}2/5$ quasihole excitations matches the CF theory up to $m{=}10$ sector. For $m{=}11,12,13$ sectors, we found the highest-energy state merges with the higher branches. The second branch starts with $m{=}0$, but we cannot identify the counting because of the absence of a clear gap. The LDOS counting of the $\nu{=}2/5$ quasiparticle excitations is shown in panel (f). We can identify the counting of the lowest branch with CF theory for all $m$ sectors, although the gap for $m{=}12,13$ becomes quite small. In this case, we can also identify the counting of the second branch up to $m{=}6$, as shown in Table~\ref{table1}. The gap between the second branch and the higher ones gradually disappears after that.

We also show the LDOS spectra for $\nu{=}3/5$ in Fig.~\ref{LDOS2_5}(g) and (h). We choose the same value for $2Q$ as in Fig.~\ref{LDOS2_5}(e) and (f). Therefore, the quasihole (quasiparticle) excitations in (g) and (h) are just particle-hole conjugates of quasiparticle (quasihole) excitations in (f) and (e). The LDOS spectra are seen to be identical, except for the unimportant energy offset due to the one-body term resulting from the particle-hole conjugation. 

Finally, the LDOS spectra for $\nu{=}3/7$ are shown in Fig.~\ref{LDOS2_5}(i) and (j). The counting in the lowest branch is exactly predicted by CF theory. The counting in the second branch matches the theory up to $m{=}5$, and we can also identify $m{=}0,1$ for the third branch. Due to the limited system size accessible to the calculation, the particle side excitation has $Q^*=-1$, which creates seven holes (rather than particles) in the fourth $\Lambda$L. Due to this coincidence, the particle side LDOS spectrum is similar to the hole side excitation for the chosen finite system.

\section{Particle entanglement spectrum}
\label{PES}

Another quantity that is related to the bulk excitations of an FQH system is the particle entanglement spectrum (PES), introduced in Ref.~\cite{Sterdyniak11}. The PES is defined by dividing the system into two parts, subsystem $A$ composed of $N_A$ particles and subsystem $B$ composed of $N_B$ particles, with the total particle number $N=N_A+N_B$. The state $|\Psi\rangle$ of the system (assumed to be non-degenerate, although the generalization to the degenerate case is straightforward~\cite{Sterdyniak11}) can be decomposed as 
\begin{eqnarray}
|\Psi\rangle=\sum_ie^{-\xi_i/2}|\Psi_i^A\rangle\otimes|\Psi_i^B\rangle,
\end{eqnarray}
where $\langle \Psi_i^A|\Psi_j^A\rangle=\langle \Psi_i^B|\Psi_j^B\rangle=\delta_{ij}$. The bipartition preserves the full symmetry of the original system and the entanglement levels $\{ \xi_i \}$ of the PES spectrum can be labelled by angular momentum quantum numbers, $L$ and $L_z$. The $\xi_i$ and $|\Psi_i^A\rangle$ can be regarded as the eigenvalues and eigenvectors, respectively, of the ``particle entanglement Hamiltonian", $H=-\ln(\rho_A)$, where $\rho_A$ is the reduced density matrix for subsystem $A$.

PES turns out to be closely related to the orbital entanglement spectrum (OES) \cite{Li08, Dubail12}, whereby the system is divided into two parts by introducing a geometric/orbital partition. Therefore, the partition in the OES is a subset of the partitions in PES, and the number of levels in OES is bounded by the number of levels in PES. OES and PES correspond to the edge excitations and bulk excitations, respectively. Ref.~\cite{Chandran11} found that for a state that is uniquely defined by some clustering properties, the counting of OES and PES in the thermodynamic limit is identical, as expected from the bulk-edge correspondence. On the other hand, for finite systems, there is a range of angular momentum sectors for which the counting of OES and PES agrees for such ``clustered" states, with the range becoming larger as the system size is increased. 

Furthermore, when the state is uniquely defined by some clustering properties, the correspondence between the number of levels in PES and the allowed number of bulk quasihole excitations has been rigorously established~\cite{Chandran11}. As explained in Ref.~\cite{Chandran11}, the partition of the clustered state inherits the clustering properties, and this sets the number of bulk quasihole excitations as an upper bound of the rank of the particle entanglement matrix. The existence of an exact clustering property, assumed in this proof, is a special feature of model FQH states described by holomorphic wave functions, which also have exact parent Hamiltonians that compactly encode the clustering conditions~\cite{Read96}. For such model states, the upper bound on the PES counting is indeed saturated. On the other hand, the focus of the present paper -- Jain states projected to the LLL -- are not uniquely identified by clustering~\cite{Regnault09, Sreejith18}. Somewhat surprisingly, it has been numerically confirmed that the counting of the PES still corresponds to the number of bulk excitations for $\nu{=}2/5$ and $3/7$ Jain states~\cite{Sterdyniak11}. This result strongly suggests that the number of levels in PES is correctly predicted by the CF theory, even in the absence of exact clustering properties.

\begin{figure}[tbh]
	\includegraphics[width=0.8\columnwidth]{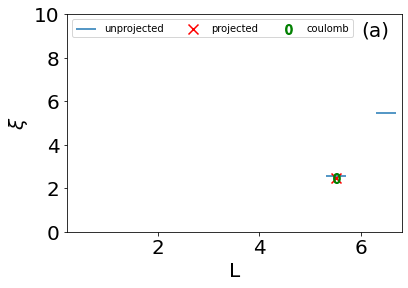}
 \includegraphics[width=0.8\columnwidth]{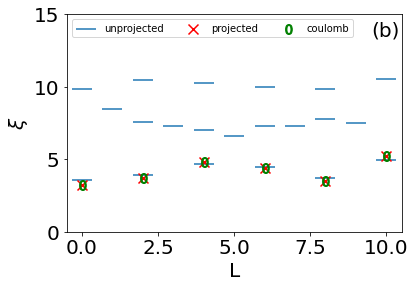} 
 \includegraphics[width=0.8\columnwidth]{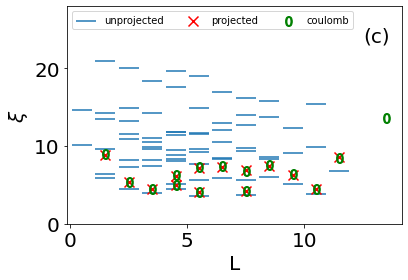} 
	\caption{The PES for unprojected Jain states (blue "$-$"), projected Jain states (red "$\times$"), and ground states of the unscreened Coulomb interaction in the LLL (green "$0$") for $N=6$, $2Q=11$ at $\nu=2/5$. The subsystem contains $N_A=1$ particles in (a), $N_A=2$ particles in (b), and $N_A=3$ particles in (c). The counting of unprojected states agrees with Eq.~\eqref{PES unpro}. Some states are annihilated by LLL projection. The low-energy part of projected Jain states and the LLL Coulomb ground states are similar, except that the level at $L=13.5$ for $N_A=3$ is only present for the Coulomb ground state.
	}
	\label{appen3}
\end{figure}

\begin{table*}
\makebox[\textwidth]{\begin{tabular}{|c|c|c|c|c|}
\hline
$N_A$&1&2&3\\ \hline
CF counting [Eq.\eqref{PES unpro}]&(0,0,0,0,0,1,1)&(2,1,3,1,3,1,3,1,3,1,2)&(2,6,7,9,11,9,7,7,5,3,3,1)\\ \hline
Unprojected Jain states&(0,0,0,0,0,1,1)&(2,1,3,1,3,1,3,1,3,1,2)&(2,6,7,9,11,9,7,7,5,3,3,1)\\ \hline
Projected Jain states&(0,0,0,0,0,1,1)&(1,0,1,0,1,0,1,0,1,0,1)&(0,1,1,1,2,2,1,2,1,1,1,1)\\ \hline
\end{tabular}}
\caption{The PES counting of unprojected and projected Jain states for $N=6$ electrons at $\nu=2/5$. The unprojected case fully agrees with Eq.~\eqref{PES unpro}.\label{table2}}
\end{table*}

In this section, we illustrate how to understand the PES counting from CF theory. We first consider the unprojected Jain states at $\nu=n/(2pn+1)$. These states live in the Hilbert space formed by the lowest $n$ LLs. When the system is divided into two parts, one with $N_A$ particles and the other with $N-N_A$ particles (and $N_A\leq N/2$), the effective field for the $N_A$ particles is $2Q^*=2Q-2p(N_A-1)$. To count the number of hole excitations for this state, we need to allocate $N_A$ particles in the $n$ $\Lambda$Ls. Take the unprojected Jain states at $\nu=2/(4p\pm 1)$ as examples ($p=1,~+$ sign for $\nu=2/5$ and $p=1,~-$ sign for $\nu=2/3$), the CF theory predicts the number of PES levels to be
\ba
\label{PES unpro}
\nonumber && (L_z,N_A,2Q^*+1)+(L_z,N_A,2Q^*+3) \\ && \nonumber +\sum_{n=1}^{N_A-1}\sum_m(m,n,2Q^*+1) \times (L_z-m,N_A-n,2Q^*+3)\\
\ea
where $Q^*=Q-p(N_A-1)$ is the effective number of flux quanta. In Eq.~\eqref{PES unpro}, the first term is the number of states with $N_A$ particles in the lowest $\Lambda$L, the second term is the number of states with $N_A$ particles in the second $\Lambda$L, and the last term sum over all states which have particles in both $\Lambda$Ls. 
We found that for all the tested cases ($N=4,6$ for $\nu=2/5$ fermionic Jain states and $N=4,6,8,10$ for $\nu=2/3$ bosonic Jain states) the PES counting agrees with the above counting. The unprojected Jain states (bosonic at $\nu{=}2/3$, fermionic at $\nu{=}2/5$) were obtained by diagonalizing the corresponding Trugman-Kivelson parent Hamiltonian in two LLs with cyclotron energy set to zero~\cite{Trugman85, Rezayi91}.

On the other hand, the more commonly studied \emph{projected} Jain states have different counting as a result of the exclusion rules mentioned in Sec.~\ref{CF}. The higher-LL part of the Hilbert space is annihilated by the LLL projection and, as a result, linear dependence is induced in the CF basis. In other words, the overlap matrix of LLL-projected CF states no longer necessarily has full rank. This can be viewed as a ``nonperturbative effect" of the CF interaction~\cite{Balram13}: the CF-CF interactions push the energies of some states to infinity after LLL projection, such that these states effectively disappear from the spectrum. The singular-value-decomposition of projected states can be written as
\be
\mathcal{P}_{\rm LLL}|\Psi\rangle=\sum_i e^{-\xi_i/2}\mathcal{P}_{\rm LLL}|u_i\rangle\otimes|v_i\rangle
\ee
On the right-hand side, $\mathcal{P}_{\rm LLL}|u_i\rangle\otimes|v_i\rangle$ corresponds to the hole excitation after projection. Since these states become linearly dependent, the number of levels in PES is also reduced. The number of levels in PES for projected states should be equal to the number of linearly independent hole excitations, the computation of the latter is discussed in Ref.~\cite{Balram13}.

In Fig.~\ref{appen3}, we compare the PES for projected and unprojected Jain states at $\nu=2/5$ for $N=6$ and show these counting in Table~\ref{table2}. Bandyopadhyay \emph{et al.}~\cite{Bandyopadhyay20} have recently proposed a set of parent Hamiltonians which have the unprojected Jain states at $\nu=n/(2pn+1)$ as highest-density zero modes. Based on this, we expect the na\"ive CF counting with $n$ $\Lambda$Ls will work for general unprojected Jain states at $\nu=n/(2pn+1)$, although we have not explicitly tested cases with $n{>}2$. On the other hand, there are fewer levels in the PES of projected Jain states as a result of the exclusion rules discussed above. In the case of $N_A=1$, $2Q^*=2Q-2p(N_A-1)=2Q$. According to the non-interacting CF picture, we now put one CF in the lowest two $\Lambda$Ls. Therefore, na\"ively we expect two states in PES, one in the lowest $\Lambda$L with $L=Q^*$ and the other in the second $\Lambda$L with $L=Q^*+1$. As Fig.~\ref{appen3}(a) shows, there is only one state with $L=Q^*$ for the projected Jain state. The other state should not exist at all, because its angular momentum exceeds the maximum angular momentum in the LLL. This does not imply the CF picture no longer works here. The $L=Q^*+1$ state gets annihilated by the LLL projection in the CF formalism because it resides fully in the second LL.

For the $N_A=2$ case, there are two configurations for the highest $L=2Q^*+1$ case: there is one particle at $L_z=Q^*+1$ in the second $\Lambda$L and the other one can be at $L_z=Q^*$, either in the second $\Lambda$L or the lowest $\Lambda$L. Thus, there are two levels expected at $L=10$ and indeed, we see two levels for the unprojected state in Fig.~\ref{appen3}(b). However, we only see one level for the projected state. The reason is similar to the explanation of the $N_A=1$ case. After the LLL projection, the two states at $L_z=10$ are not linearly independent -- there is only one independent state. For the same reason, the PES level pattern for $N_A=2$ is $1,0,1,0,1,0,1,0,1,0,1$ for projected states, which is actually the full dimension of the lowest LL. 

The PES levels of the projected states closely coincide with the lower part of the PES levels of the unprojected states, which is expected as they lie in the same topological phase \cite{Anand22}. In Fig.~\ref{appen3}, we also include the PES of the LLL unscreened Coulomb ground state, which is very similar to the PES of the projected Jain state. However, the full counting of the PES of the LLL Coulomb ground state is given by $(L_z, N_A,2Q)$. For $N_A=1,2$, this coincides with the counting of the projected Jain state. For $N_A\geq 3$, only the lower part of the LLL Coulomb ground state is expected to be similar to the projected Jain state.

One might wonder why we do not need to consider the exclusion rules for the lowest branch of LDOS counting discussed in Sec.~\ref{LDOS}. In that case, we only have quasiparticles in the lowest unoccupied $\Lambda$L or quasiholes in the highest occupied $\Lambda$L. As shown in Ref.~\cite{Balram13}, the CF-CF interactions do not eliminate any states in such cases. However, more generally, when there is more than one $\Lambda$L that is partially occupied, the strong effect of CF interaction will push the energies of some states to infinity or, in other words, some states will be annihilated by the LLL projection in the CF formalism. For the PES counting of Jain states, we must take into account such nonperturbative effects since there are two or more partially occupied $\Lambda$Ls. A detailed discussion of this strong effect or exclusion rule is given in Ref.~\cite{Balram13}. In certain cases, the simple exclusion rules specified in Ref.~\cite{Balram13} need to be supplemented with additional exclusion rules. To get the accurate CF counting with the CF-CF interaction included, one can use the technique of CF diagonalization \cite{Jain97, Mandal02} to find out the precise number of independent states. Notably, the exclusion principle plays no role in the counting of PES at $\nu{=}1/3$, since only the lowest $\Lambda$L is involved.

\begin{figure}[tb]
\includegraphics[width=0.49\columnwidth]{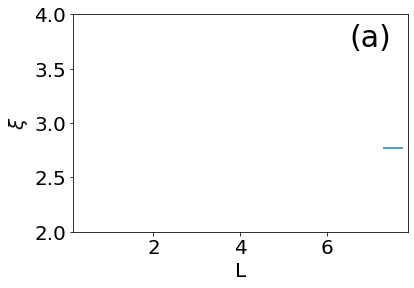}
\includegraphics[width=0.49\columnwidth]{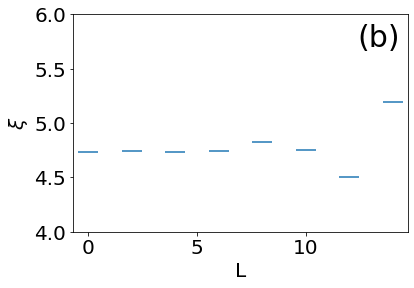}
\includegraphics[width=0.49\columnwidth]{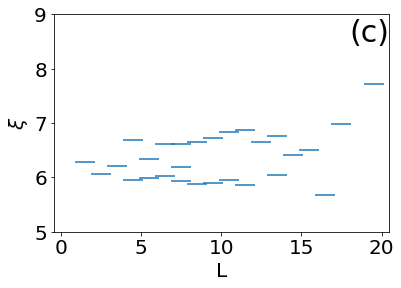}
\includegraphics[width=0.49\columnwidth]{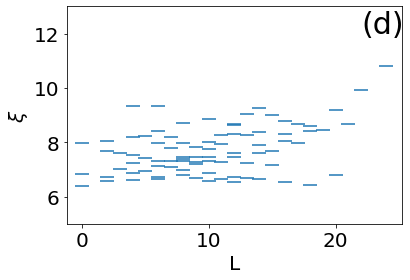}
\includegraphics[width=0.49\columnwidth]{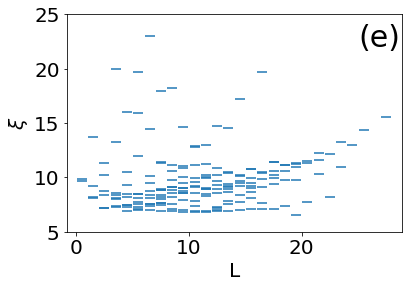}
\caption{ 
The PES for $N=10$ electrons and $2Q=15$ flux quanta corresponding to $\nu=2/3$ Jain state. The subsystem contains $N_A=1,2,3,4,5$ electrons in panels (a), (b), (c), (d), and (e), respectively. In this case, the counting is exactly given by the number of states in the full Hilbert space, i.e., $N_A$ particles in $2Q+1$ orbitals.
\label{PES2_3}}
\end{figure}

We also studied the PES of the projected $\nu=n/(2pn-1)$ Jain states. The ground states at these fillings have reversed vortex attachment, which means $Q^*<0$. On the other hand, for the subsystem $A$ with $N_A\leq N/2$, the effective magnetic field is reversed back to $Q_A^*>0$. Therefore, the $\Lambda$L structure completely changes in going from the ground states to the excited states in PES, and as a result CF theory cannot give a prediction on the counting. We illustrate this using the $\nu=2/3$ state, whose PES is presented in Fig.~\ref{PES2_3}. We find that, for each $N_A$ and each $L$, the counting of PES is exactly given by the number of excited states in the lowest LL (i.e., \emph{not} $\Lambda$L). This far exceeds the number of states in the lowest two $\Lambda$Ls if we apply the CF theory. Thus, the CF theory is not useful in predicting the PES counting for the reversed-vortex attached states. This does not mean the CF theory is inconsistent with PES, rather the CF theory does not predict a PES counting for 2/3. We expect it to be generally true for the $n/(2pn-1)$ Jain states that the counting is simply given by the full counting in the LLL and the $\Lambda$L structure is irrelevant. 

\begin{figure}[tb]
	\includegraphics[width=0.35\textwidth]{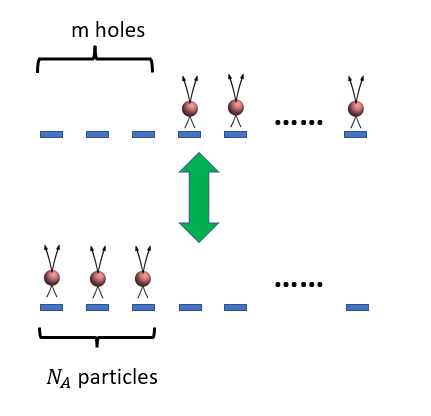}
	\caption{Matching the LDOS and PES countings for the Laughlin states. There are $m$ holes in the lowest $\Lambda$L for LDOS and $N_A=2p+1$ particles in the lowest $\Lambda$L for PES, and the total numbers of orbitals are the same $2\tilde{Q}-2p(\tilde{N}-2)+1=2Q-2p(N_A-1)+1$.}
	\label{LDOS-PES-matching}
\end{figure}

\begin{figure*}[ht]
	\includegraphics[width=1.0\textwidth]{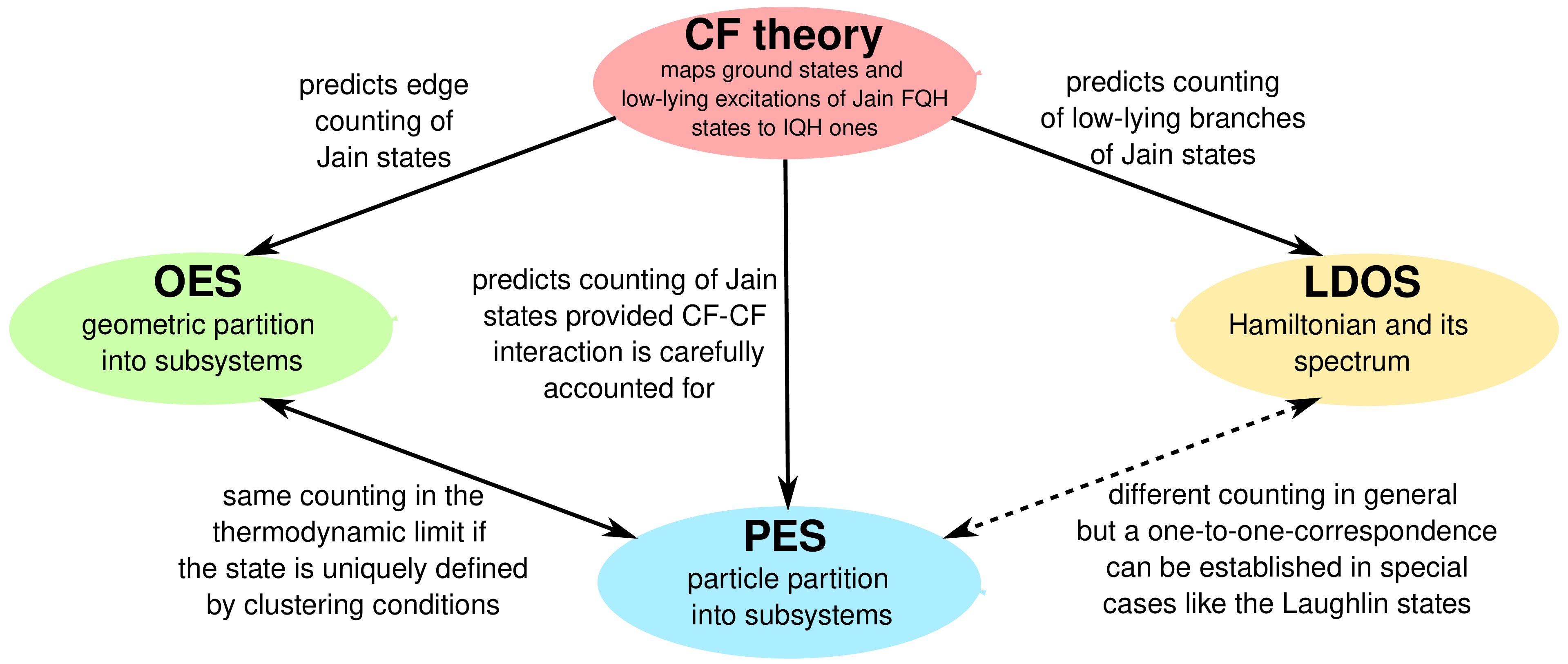}
	\caption{Summary of the main results. We study two quantities that both describe the bulk excitations of FQH systems, LDOS and PES, using CF theory. The CF theory predicts the counting of LDOS if the electron-electron interaction opens up gaps between different branches of CF excitations. The PES counting can also be predicted from CF theory, after carefully incorporating the residual interaction between CFs. Furthermore, PES has previously been related to OES~\cite{Chandran11}, which describes the edge excitations, if the model state is uniquely defined by clustering properties.	}
	\label{fig1}
\end{figure*}

\section{The relation between LDOS and particle entanglement}
\label{discussion}

Since both the LDOS and PES count the bulk quasihole excitations, one might na\"ively expect that a one-to-one correspondence could be established between them. However, this turns out to not be true in general. There is a crucial difference in how LDOS and PES each measure the bulk quasihole excitations. The LDOS creates one hole (equivalent to a few quasiholes) on top of the ground state by removing one electron, while the PES creates $N-N_A$ holes on top of the ground state, with the restriction $N_A\leq N/2$, where $N$ is the electron number for the ground state. When $N_A>N/2$, the PES is the same as $N_A'=N-N_A$, which equals the number of excitations creating $N_A$ holes. Therefore, the PES always describes the creation of $\max\{N_A, N-N_A\}$ holes, which means the number of holes is always more than half of the electron number in the ground state. 

The difference between LDOS and PES becomes even more obvious in the thermodynamic limit. In the thermodynamic limit, the counting of LDOS is equivalent to the number of states of creating one hole on the ground state while the counting of PES is equivalent to the number of states obtained by creating a finite number of particles on top of the vacuum state. The latter can also be viewed as creating an infinite number of holes in the ground state, which is very different from creating one hole in the ground state. The counting for the PES is still well-defined by counting the configurations of a finite number of particles allowed by the clustering property, although this is not equivalent to the configurations of a finite number of holes allowed by the clustering property. 

Nevertheless, while a one-to-one correspondence between LDOS and PES may not exist in general, a correspondence can be established in special cases such as the Laughlin states through CF theory. Suppose the ground state has $\tilde{N}$ electrons, and $2\tilde{Q}=(2p+1)\tilde{N}-(2p+1)$ flux quanta, which corresponds to the $\nu=1/(2p+1)$ Laughlin state. For the LDOS, we remove one electron from the system. The counting is the same as the number of zero modes for the Haldane pseudopotential interaction $\sum_{l=1}^pV_{2l-1}$ with $\tilde{N}-1$ electrons in $2\tilde{Q}$ flux quanta. For the PES counting, we assume there are $N$ electrons divided into two subsystems, with $N_A\leq N_B$. The number of flux quanta is $2Q=(2p+1)N-(2p+1)$. The counting equals the number of zero modes for the Haldane pseudopotential interaction $\sum_{l=1}^pV_{2l-1}$ with $N_A$ electrons in $2Q$ flux quanta.

From CF theory, we know the exact number of zero modes. On the LDOS side, it is the number of states with $2p+1$ holes in the $2\tilde{Q}+1-2p(\tilde{N}-2)$ CF orbitals. On the PES side, it is the number of states with $N_A$ particles in $2Q+1-2p(N_A-1)$ CF orbitals. There are two ways in which parameters can be chosen such that the LDOS and PES have the same counting. The first way to match the number of zero modes is by imposing $Q=\tilde{Q}$ and $N_A=\tilde{N}-1$, but this does not work as it violates our constraint $N_A\leq N/2$. However, there is a second option: we can choose 
the number of holes in LDOS to be the same as the number of particles $N_A$ in PES, and also make the effective magnetic fields the same,
as illustrated in Fig.~\ref{LDOS-PES-matching}. This corresponds to
\be
\label{LDOS-PES}
\begin{cases}
2p+1=N_A,\\
2\tilde{Q}-2p(\tilde{N}-2)=2Q-2p(N_A-1),\\
N_A\leq N/2.\\
\end{cases}
\ee
The solution to the above equations is
\be
\begin{cases}
N_A=2p+1,\\
N=\frac{1}{2p+1}\tilde{N}+2p+1-\frac{1}{2p+1},\\
\tilde{N}\geq (2p+1)^2+1.\\
\end{cases}
\ee
Because of the inequality constraint, $\tilde{N}$ must be no less than $10$. All choices of $(2Q,N,N_A|2\tilde{Q},\tilde{N})$ given by the above equations have the same counting for LDOS spectrum and PES spectrum, as we have confirmed numerically for $(15,6,3|27,10)$, $(18,7,3|36,13)$, and $(21,8,3|45,16)$. 

It is natural to ask whether a similar relationship like Eq.~\eqref{LDOS-PES} can be established for general Jain states. There are several obstacles to doing so. To match the full counting of LDOS and PES, a parent Hamiltonian for the Jain states is needed to compute the LDOS. To the best of our knowledge, such Hamiltonians do not exist for the projected Jain states~\cite{Rezayi91, Sreejith18, Bandyopadhyay20}. Without a parent Hamiltonian, the best one can do is to identify the lower-energy branches of LDOS with PES. However, the PES generally does not exhibit easily identifiable branches organized hierarchically, i.e., a higher energy state does not correspond to a higher entanglement energy state. Therefore, we expect in general it would be challenging to relate the LDOS and PES level counting.

\section{Conclusion}
\label{conclusion}

The main conclusions of this paper are summarized in Fig.~\ref{fig1}. In this work, we studied two quantities, LDOS and PES, that both describe the bulk excitations of FQH systems, and we showed that both quantities can be understood using CF theory for Jain states. We demonstrated how to predict the LDOS spectrum counting from the CF theory using various Jain states as examples. Numerical simulations based on the screened Coulomb interaction show that the lowest branch of LDOS spectra agrees well with theoretical predictions for several angular momentum sectors, while in some cases the second or even the third branch counting is also identifiable for the first few angular momentum sectors. Moreover, we studied the PES counting from the CF theory. We found the PES counting for Jain states at $\nu{=}n/(2pn+1)$ to be accurately predicted by CF theory. For Jain states at $\nu{=}n/(2pn-1)$ with reversed vortex attachment, the counting is given by the number of states in the full lowest LL, which is beyond the subspace described by CF theory. We also discussed the relationship and differences between LDOS and PES. While a general one-to-one mapping between them is likely impossible to establish, we showed that in special cases, such as the Laughlin states, one can indeed map the LDOS and PES counting onto each other with an appropriate choice of system sizes. 

Our results are expected to apply to other material systems that realize FQH phases. Apart from GaAs, the Jain sequence of FQH states is also known to occur in monolayer graphene~\cite{Bolotin09, Du09, Amet15, Balram15c, Kim19}, bilayer graphene~\cite{Diankov16, Zibrov17, Li17, Huang21, Balram21b} and transition metal dichalcogenides such as WSe$_2$~\cite{Shi19}. Moreover, members of the $n/(4n\pm 1)$ Jain sequence observed in the second LL of GaAs~\cite{Kumar10} are likely analogous to their LLL counterparts~\cite{Balram20a} and thus are also expected to lend themselves to a description in terms of CFs.

Several questions remain to be answered. It is not clear under what general conditions the number of PES levels equals the number of bulk quasihole excitations. For FQH states with known parent Hamiltonians, such as the Laughlin states~\cite{Laughlin83}, Moore-Read states~\cite{Moore91}, and the Read-Rezayi $\mathbb{Z}_3$ states~\cite{Read99}, we have numerically confirmed that the number of levels in PES exactly matches the number of zero modes of the parent Hamiltonian for $N_A$ particles in the same total magnetic field (the first two cases have also been numerically confirmed by Sterdyniak~\emph{et al.}~\cite{Sterdyniak11}). For the Jain states considered in this work, the number of bulk excitations is calculated using CF theory rather than counting the zero modes of parent Hamiltonian. Bandyopadhyay~\emph{et al.}~\cite{Bandyopadhyay20} have recently constructed a parent Hamiltonian for the $\nu=n/(2pn+1)$ unprojected Jain states. The correspondence between PES and bulk excitations for projected Jain states, which do not have any known parent Hamiltonian, could be understood if the equivalence between the number of excitations and PES levels survives the LLL projection. However, it is not clear why that must be the case. Moreover, for a generic state with no parent Hamiltonian, it is not clear how to relate the number of excitations and levels in PES. 

An interesting question is how to predict the LDOS and PES countings for states obtained from parton theory~\cite{Jain89b}, which generalizes the CF construction by expressing FQH states as products of IQH states. Several non-CF parton states have recently been shown to be potential candidates for describing the ground and excited FQH states in multilayer graphene, second LL, and in wide quantum wells~\cite{Wu17, Balram18, Balram18a, Balram19, Faugno19, Balram21d, Dora22}. Many of these parton states support non-Abelian excitations~\cite{Wen91} and it is as yet unclear how the branch structure and the counting in each branch works out for these states. Moreover, parent Hamiltonians are not known for generic unprojected parton states. It would be interesting to check whether the number of levels in the PES spectra of parton states matches the number of their bulk excitations.

\section{ Acknowledgments} 
Z.P. would like to thank Ali Yazdani, Roger Mong and Michael Zaletel for previous collaboration on the same topic that motivated this project. S. P. and Z. P. acknowledge support by the Leverhulme Trust Research Leadership Award RL-2019-015 and by EPSRC grant EP/R020612/1. A. C. B. acknowledges the Science and Engineering Research Board (SERB) of the Department of Science and Technology (DST) for financial support through the Start-up Grant SRG/2020/000154. A. C. B. and Z. P. thank the Royal Society International Exchanges Award IES$\backslash$R2$\backslash$202052 for funding support. Some of the numerical calculations reported in this work were carried out on the Nandadevi supercomputer, which is maintained and supported by the Institute of Mathematical Science’s High-Performance Computing Center. Statement of compliance with EPSRC policy framework on research data: This publication is theoretical work that does not require supporting research data.

\begin{appendix}

\section{Optimizing the parameters for LDOS spectrum}
\label{parameters}

\begin{figure}[tbh]
	\includegraphics[width=0.35\textwidth]{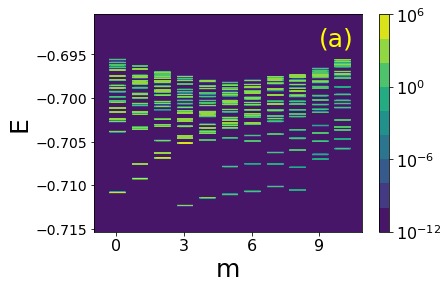}
 \includegraphics[width=0.35\textwidth]{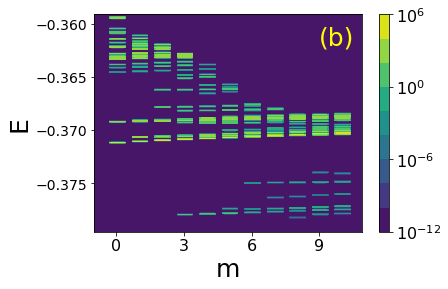} 
	\caption{The LDOS spectrum for $N=9$, $2Q=21$, corresponding to $\nu{=}1/3$ particle excitations. The screening distances and impurity distances are $d_g=7\ell$, $d_i=2.1\ell$ for (a) and $d_g=3\ell$, $d_i=2.9\ell$ for (b). We cannot identify the counting in (a) with the counting expected from CF theory for $m>4$, while there is a clear gap in (b) which allows us to match the countings for all $m$ sectors. In both cases, the vacuum state $\Omega$ has a large overlap with the Laughlin state, i.e., $|\langle \Omega|\Psi^{\rm Laughlin}_{1/3}\rangle|^2=0.9536$ for (a) and $0.9906$ for (b). 
	}
	\label{appen1}
\end{figure}

In this Appendix, we explain how we determine the optimal parameters for identifying the LDOS spectrum counting. Even if the vacuum state $|\Omega\rangle$ has a large overlap with a model state such as the Laughlin state, $|\Psi^{\rm Laughlin}_{1/3}\rangle$, the LDOS excitation spectrum can show big deviations from the Laughlin excitation spectrum, depending on the details of the interactions and impurity potential. The parameters $d_i$ and $d_g$ defined in Eqs.~(\ref{interaction})-(\ref{potential}) affect the gaps between different branches of the LDOS spectrum, as illustrated in Fig.~\ref{appen1}. Even with relatively small changes in $d_i$ and $d_g$, the spectrum can look very different. We would like to optimize the parameters so that the counting in the lowest branch can be easily identified by visual inspection. To achieve this aim, we perform a systematic scan of $d_g$ and $d_i$. We set the standard for ``good parameters" as follows. We first set a critical $m$ value for which the gap is sensitive to the parameters for each filling factor respectively. This is usually the third or fourth number counting from the starting $m$ since the gaps for smaller $m$ are always pretty big and the gaps for bigger $m$ are always very small. If $n$ states are expected to occur in the lowest branch for the $m$th sector, we use 
\be
\gamma\equiv\frac{(n-1)(E_n-E_{n-1})}{E_{n-1}-E_0}
\label{gamma}
\ee
to quantify the gap which separates the lowest branch from higher branches in the $m$th sector. This is the ratio of the gap between the $n$th and $(n-1)$th state and the average value of the gaps between the $0$th state and $n-1$th state. 

\begin{figure}[tb]
 \includegraphics[width=0.45\columnwidth]{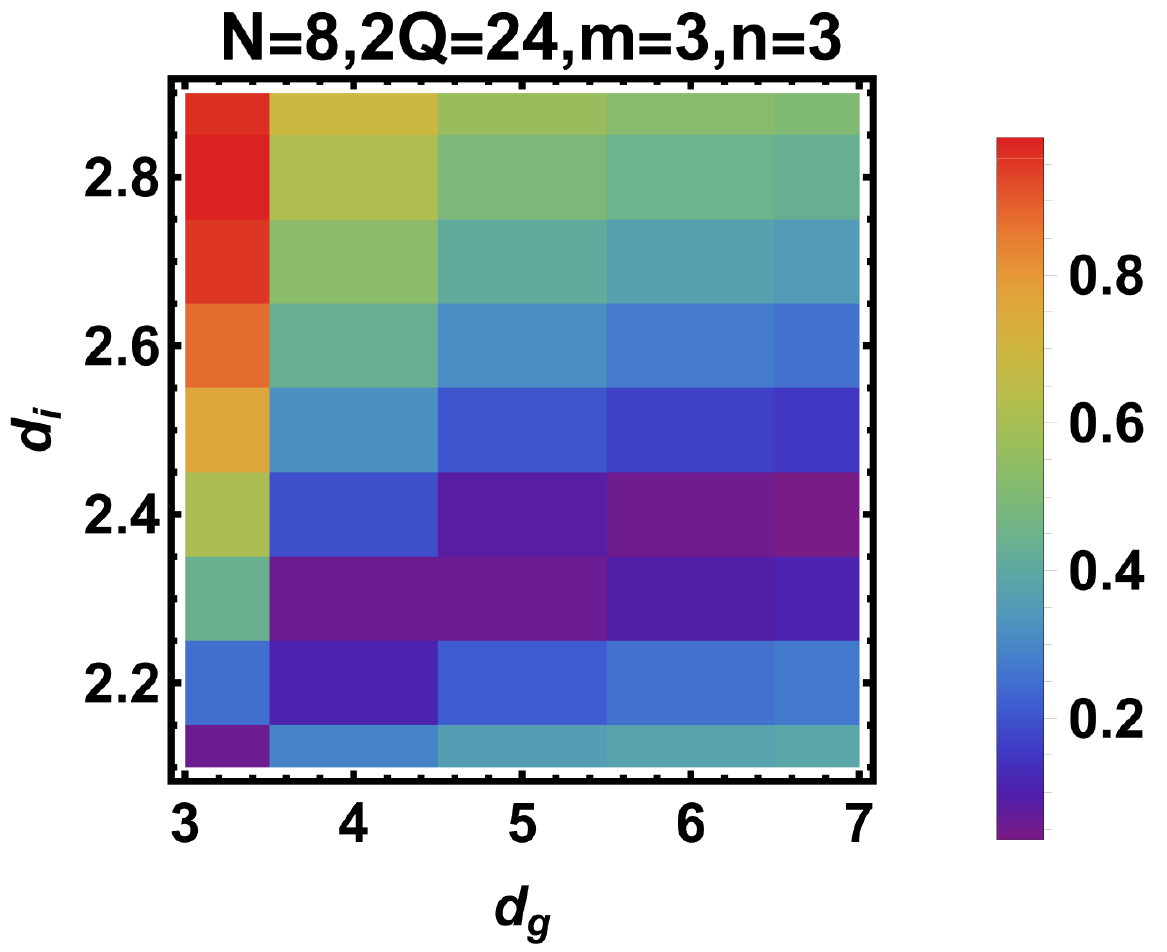}
 \includegraphics[width=0.45\columnwidth]{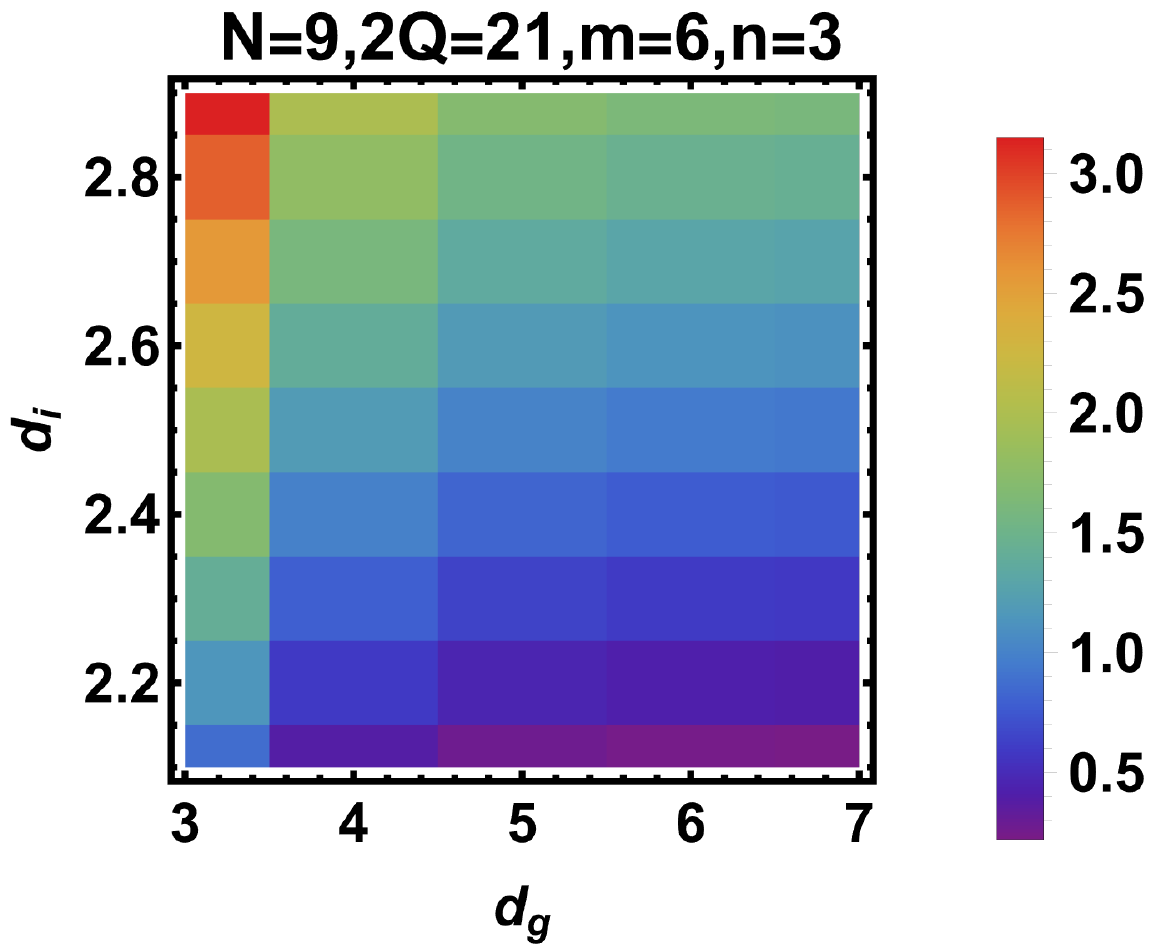} 
 \includegraphics[width=0.45\columnwidth]{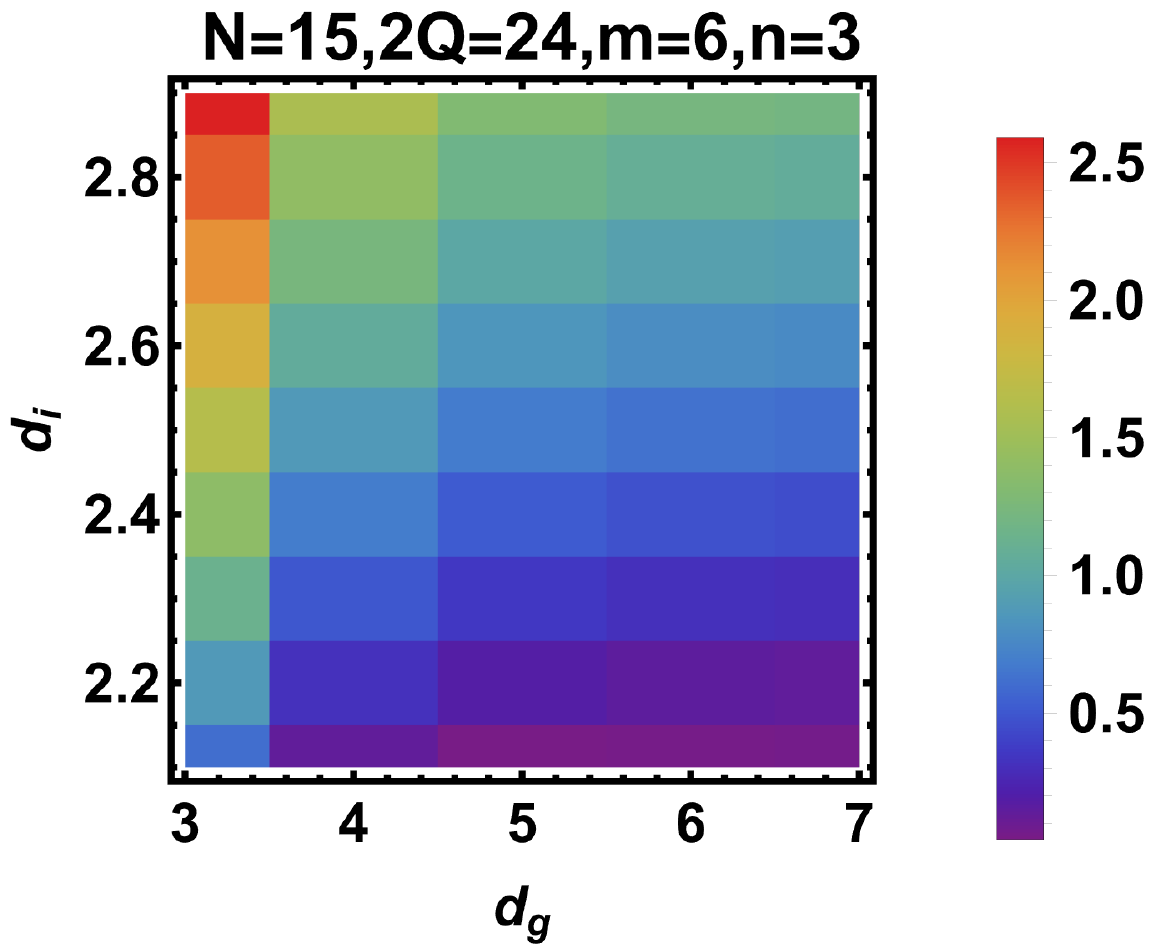} 
 \includegraphics[width=0.45\columnwidth]{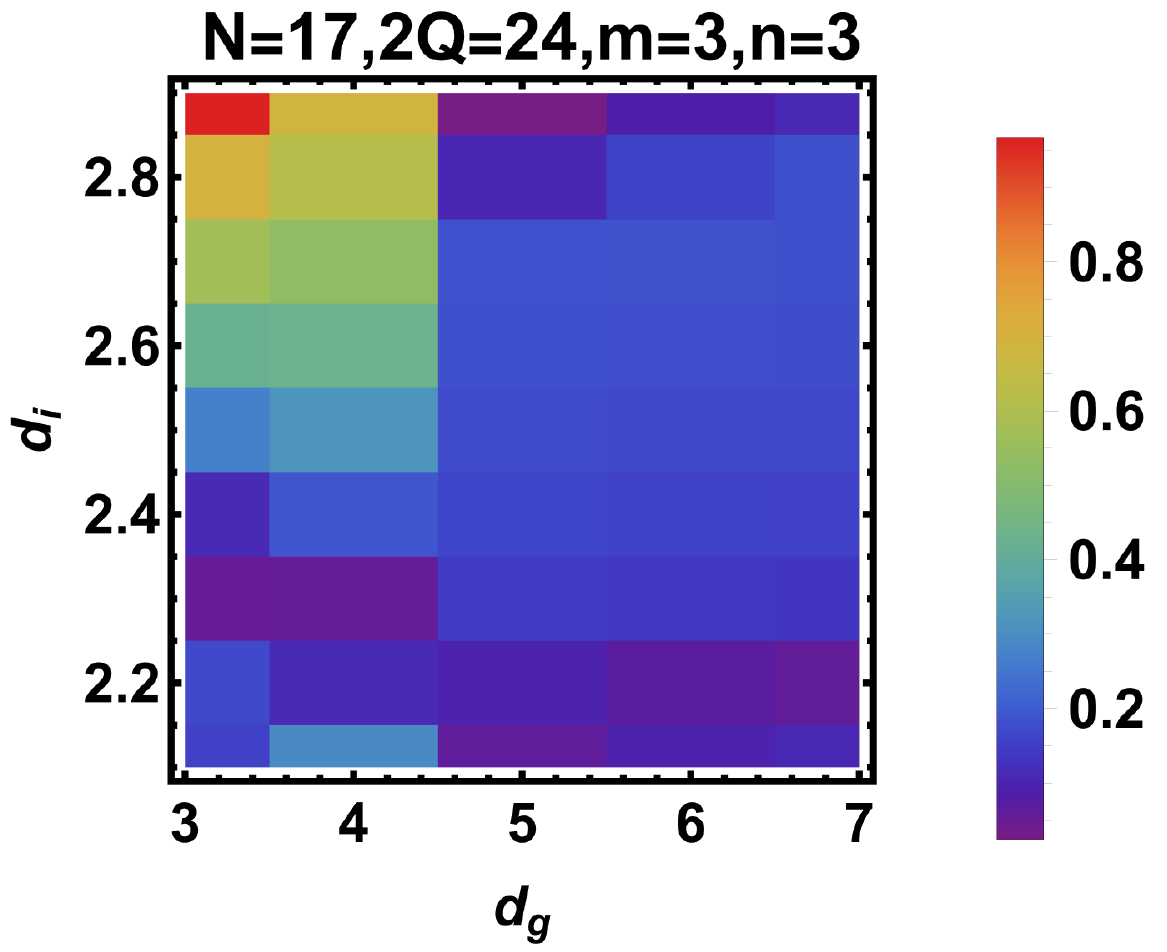} 
	\caption{Color scale shows the value of $\gamma$, defined in Eq.~\eqref{gamma}, in the parameter space $d_i-d_g$ for several filling factors. $N=8, 2Q=24$ is a $\nu{=}1/3$ hole excitation state. $N=9, 2Q=21$ is a $\nu{=}1/3$ particle excitation state, while it is also a $\nu{=}2/5$ hole excitation state (CF theory predicts the same counting for both states). $N=15, 2Q=24$ is a $\nu{=}2/3$ hole excitation state. $N=17,2Q=24$ is a $\nu{=}2/3$ particle excitation state. All these results suggest that the optimal parameters are around $d_i=2.9\ell$, $d_g=3\ell$. The squares of overlaps between the original states and the Laughlin state $|\langle \Omega|\Psi^{\rm Laughlin}_{1/3}\rangle|^2>0.9$ for all states.
	}
	\label{appen2}
\end{figure}

In Fig.~\ref{appen2}, the ratio $\gamma$ is computed across the parameter space for filling factors $\nu{=}1/3$ and $2/3$. In these examples, the squared overlap between the exact ground state and the model Laughlin wave function is always larger than 0.9, yet the value of $\gamma$ is quite sensitive to the parameters. These results suggest that the optimal parameters are around $d_i=2.9\ell$, $d_g=3\ell$, i.e., the LDOS counting is easiest to identify when the screening distance is small and the impurity distance is close to the screening distance. We use such parameters for our calculations in the main text.

\end{appendix}

\end{document}